\documentclass[a4paper,fleqn,usenatbib]{mnras}

\usepackage{newtxtext,newtxmath}

\usepackage[T1]{fontenc}
\usepackage{ae,aecompl}

\usepackage{graphicx}	
\usepackage{amsmath}	
\usepackage{amssymb}	
\usepackage{multirow}   
\usepackage{longtable}  
\usepackage{graphicx}
\usepackage{subfig}

\title[Stellar activity analysis of Barnard's Star]{Stellar activity analysis of Barnard's Star: Very slow rotation and evidence for long-term activity cycle}

\author[B. Toledo-Padr\'on et al.]{
B. Toledo-Padr\'on,$^{1,2}$ \thanks{E-mail: btoledo@iac.es}
J. I. Gonz\'alez Hern\'andez,$^{1,2}$
C. Rodr\'iguez-L\'opez,$^{3}$\newauthor
A. Su\'arez Mascare\~no,$^{4}$
R. Rebolo,$^{1,2,5}$
R. P. Butler,$^{6}$
I. Ribas,$^{7,8}$
G. Anglada-Escud\'e,$^{3,9}$\newauthor
E. N. Johnson,$^{10}$
A. Reiners,$^{10}$
J. A. Caballero,$^{11}$
A. Quirrenbach,$^{12}$
P. J. Amado,$^{3}$ \newauthor
V. J. S. B\'ejar,$^{1,2}$
J. C. Morales,$^{7,8}$
M. Perger,$^{7,8}$
S. V. Jeffers,$^{10}$
S. Vogt,$^{13}$
J. Teske,$^{6}$ \newauthor
S. Shectman,$^{14}$
J. Crane,$^{14}$
M. D\'iaz,$^{14,15}$
P. Arriagada,$^{6}$
B. Holden,$^{13}$
J. Burt,$^{16}$ \newauthor
E. Rodr\'iguez,$^{3}$
E. Herrero,$^{7,8}$
F. Murgas,$^{1,2}$
E. Pall\'e,$^{1,2}$
N. Morales,$^{3}$ \newauthor
M. J. L\'opez-Gonz\'alez,$^{3}$
E. D\'iez Alonso,$^{17}$
M. Tuomi,$^{18}$
M. Kiraga,$^{19}$
S. G. Engle,$^{20}$ \newauthor
E. F. Guinan,$^{20}$
J. B. P. Strachan,$^{9}$
F. J. Aceituno,$^{3}$
J. Aceituno,$^{21}$
V. M. Casanova,$^{3}$ \newauthor
S. Mart\'in-Ruiz,$^{3}$
D. Montes,$^{22}$
J. L. Ortiz,$^{3}$
A. Sota$,^{3}$
J. Briol,$^{23}$
L. Barbieri,$^{23}$ \newauthor
I. Cervini,$^{23}$
M. Deldem,$^{23}$
F. Dubois,$^{23,24}$
F. -J. Hambsch,$^{23,25}$
B. Harris,$^{23}$
C. Kotnik,$^{23}$ \newauthor
L. Logie,$^{23,24}$
J. Lopez,$^{23}$
M. McNeely,$^{23}$
Y. Ogmen,$^{23}$
L. P\'erez,$^{23}$
S. Rau,$^{23,24}$ \newauthor
D. Rodr\'iguez,$^{23}$
F. S. Urquijo,$^{23}$
and S. Vanaverbeke$^{23,24}$
}

\date{Accepted XXX. Received YYY; in original form ZZZ}

\pubyear{2019}

\begin{document}
\label{firstpage}
\pagerange{\pageref{firstpage}--\pageref{lastpage}}
\maketitle

\newpage

\begin{abstract}
The search for Earth-like planets around late-type stars using ultra-stable spectrographs requires a very precise characterization of the stellar activity and the magnetic cycle of the star, since these phenomena induce radial velocity (RV) signals that can be misinterpreted as planetary signals. Among the nearby stars, we have selected Barnard's Star (Gl 699) to carry out a characterization of these phenomena using a set of spectroscopic data that covers about 14.5 years and comes from seven different spectrographs: HARPS, HARPS-N, CARMENES, HIRES, UVES, APF, and PFS;  and a set of photometric data that covers about 15.1 years and comes from four different photometric sources: ASAS, FCAPT-RCT, AAVSO, and SNO. We have measured different chromospheric activity indicators (H$\alpha$, Ca~{\sc II}~HK and Na I D), as well as the FWHM of the cross-correlation function computed for a sub-set of the spectroscopic data. The analysis of Generalized Lomb-Scargle periodograms of the time series of different activity indicators reveals that the rotation period of the star is 145 $\pm$ 15 days, consistent with the expected rotation period according to the low activity level of the star and previous claims. The upper limit of the predicted activity-induced RV signal corresponding to this rotation period is about 1 m/s. We also find evidence of a long-term cycle of 10 $\pm$ 2 years that is consistent with previous estimates of magnetic cycles from photometric time series in other M stars of similar activity levels. The available photometric data of the star also support the detection of both the long-term and the rotation signals.
\end{abstract}



\section{Introduction}

Since the discovery of the first extrasolar planet in 1992 \citep{Wolszczan1992} and the detection of the first exoplanet orbiting a solar-type star \citep{Mayor1995}, 3884 exoplanets have been detected using different techniques \footnote{source: \url{http://www.exoplanet.eu}}. One of the most commonly used methods is the radial velocity (RV) technique, which has been applied to find 773 extrasolar planets around 576 stars. The majority of these stars are G or K type (with a percentage of 42\% and 33\% respectively), and only 49 of them are M-dwarfs (8\% of the total sample), the first one detected in 2001 around GJ 876 \citep{Marcy2001}. The search for Earth-like planets around these M type stars takes advantage of having greater amplitudes in the RV planetary signals due to the low mass of their parent star. Also, this type of stars is the most common stellar type in the Milky Way \citep{Chabrier2000}. However, stellar activity in M-dwarfs can produce signals with periods commensurate with the ``habitable zones'' around these stars \citep{Vanderburg2016,Newton2016b}, where liquid water could potentially exist on the surface of a planet. Distinguishing whether signals arise from orbiting planets or stellar activity can be challenging. The signals produced by stellar activity are on the timescale of the rotation period of the star, but we also have to take care of the long-period signals associated with Doppler shifts caused by the magnetic cycle of the star \citep{Dravins1985,Campbell1988} already reported around M stars \citep{Gomes2012,Robertson2013,Díez-Alonso2019}. Photon noise of the measurements is a key selection criteria of stellar samples in RV search programs. The high SNR of nearby stars makes them very interesting targets for low mass exoplanets searches. Among the nearby stars, we have selected the closest single M-dwarf to the Solar System: Barnard's Star (Gl 699).

Barnard's Star is well known for being the second closest stellar system to the Sun. Located at a distance of 1.8 parsecs \citep{Brown2018}, and with an age between 7 and 10 Gyr \citep{Ribas2018}, Gl 699 is the star with the highest proper motion known to date \citep{Barnard1916}, which causes Doppler shifts due to secular acceleration \citep{Stumpff1985,Kurster2003} that needs to be taken into account for exoplanet searches through RV. It also presents a low X-ray luminosity, which indicates a low level of current magnetic activity \citep{Vaiana1981,Hunsch1999,Marino2000}. This reduces the effects of spots and plages in the spectral line profiles \citep{Lovis2011}. The most important properties of this star are shown in Table \ref{tab:GJ699_Propiedades}.

\begin{table}
\centering
	\caption{Stellar properties of Barnard's Star.}
	\label{tab:GJ699_Propiedades}
	\begin{tabular}{lccr} 
		\hline
		Parameter & Gl 699 & Ref.\\
		\hline
		RA (J2000)                                  & 17:57:48.50           & [1] \\
		DEC (J2000)                                 & +04:41:36.11          & [1] \\
		$\mu_{\alpha}$ cos $\delta$ (mas yr$^{-1}$) & -802.8 $\pm$ 0.6      & [1] \\
		$\mu_{\delta}$ (mas yr$^{-1}$)              & +10362.5 $\pm$ 0.4    & [1] \\
		Distance [pc]                               & 1.8266 $\pm$ 0.0001   & [1] \\
		$m_{\rm B}$                                 & 11.24                 & [2] \\
		$m_{\rm V}$                                 & 9.51                  & [2] \\
		Spectral type                               & M3.5V                 & [3] \\
		\textit{T}$_{\rm eff}$ [K]                  & 3278 $\pm$ 51         & [4] \\
	    $[$Fe/H] (dex)                              & -0.12 $\pm$ 0.16      & [4] \\
		M$_{\star}$ [M$_{\odot}$]                   & 0.163 $\pm$ 0.022     & [5] \\
		R$_{\star}$ [R$_{\odot}$]                   & 0.178 $\pm$ 0.011     & [5] \\
		L$_{\star}$ [L$_{\odot}$]                   & 0.00329 $\pm$ 0.00019 & [5] \\
		log \textit{g} (cgs)                        & 5.10 $\pm$ 0.07       & [4] \\
		log (L$_{\rm x}$/L$_{\rm bol}$)             & -5.4                  & [6] \\
		\textit{v} sin \textit{i} [km s$^{-1}$]     & $<$3                  & [4] \\
		a$_{\rm sec}$ [m s$^{-1}$ yr$^{-1}$]        & 5.15 $\pm$ 0.89       & [7] \\
		log$_{10}$ ($R^{'}_{\rm HK}$)               & -5.82 $\pm$ 0.08      & [8] \\
		$P_{\rm rot}$ [days]                        & 145 $\pm$ 15          & [8] \\
		Long-term activity cycle [days]             & 3800 $\pm$ 600        & [8] \\
		\hline
	\end{tabular}
	\begin{minipage}{\columnwidth} 
{\footnotesize \textbf{References:} [1] \citet{Brown2018}; [2] \citet{Koen2010}; [3] \citet{Alonso-Floriano2015}; [4] \citet{Passegger2018}; [5] \citet{Ribas2018}; [6] \citet{Kiraga2007}; [7] \citet{Kurster2003}; [8] This work}
\end{minipage}	
\end{table}

Previous work carried out on this star \citep{Mascareño2015,Astudillo2017} has revealed a low level of chromospheric emission (log$R^{'}_{\rm HK}$=-5.86 and log$R^{'}_{\rm HK}$=-5.69, respectively), which is usually related to slow rotators. Using these two values in the relation between the rotation period and the activity level of the star predicted by \citet{Mascareño2016} gives an expected rotation period of 152 and 112 days, respectively. This range is in good agreement with the previous value of 130 days given by \citet{Benedict1998} through a photometric study using the Hubble Space Telescope. Also \citet{Mascareño2015} reported a 148.6-day rotation period obtained from a time-series analysis of spectroscopic indexes using HARPS data.

Recently, \citet{Ribas2018} reported the discovery of a super-Earth like planet orbiting Barnard's Star at an orbital period of 233 days with a minimum mass of 3.3 Earth masses. We will focus on the stellar activity and magnetic cycle characterization through a multi-spectrograph analysis of several activity indexes, complemented by a multi-instrumental analysis of photometric time-series, which leads to detect and increase the precision in the rotation period value and also to detect a long-term activity cycle in the star.

In Section \ref{sec:Data} we describe the whole dataset, both the spectroscopy and photometry used in this work. In Section \ref{sec:Method} we describe the methodology used in the analysis of each stellar activity indicator. In Section \ref{sec:Analysis} we show this analysis and the results obtained for each activity indicator. In Section \ref{sec:Discussion} we discuss the results, and we provide the conclusions of this study in Section \ref{sec:Conclusions}.

\section{Data}

\label{sec:Data} 

\subsection{Spectroscopic Dataset}

For this work we have used spectra taken with seven different spectrographs, whose main properties are shown in Table \ref{tab:Spectrographs}.

\begin{table}
\centering
	\caption{Properties of all the spectrographs used in this work.}
	\label{tab:Spectrographs}
	\begin{tabular}{lccc}
		\hline
		Spectrograph & R & $\Delta \uplambda$ [\AA] & N$_{\rm spec}$ \\ 
		\hline
		HARPS & 115 000 & 3780-6910 & 317 \\ 
		HARPS-N & 115 000 & 3830-6930 & 74 \\ 
		CARMENES & 90 000 & 5200-17100 & 192 \\ 
		HIRES & 67 000 & 3700-10000 & 179 \\ 
		UVES & 130 000 & 3000-11000 & 57 \\ 
		PFS & 80 000 & 3880-6680 & 43 \\ 
		APF & 100 000 & 3740-9700 & 95 \\ 
		\hline
	\end{tabular}
	\begin{minipage}{\columnwidth} 
{\footnotesize \textbf{Columns:} Name of the spectrograph, resolution, spectral range and number of spectra used in this work. The observation programs are listed in the acknowledgments.}
\end{minipage}	
\end{table}

HARPS (High Accuracy Radial velocity Planet Searcher) is a fiber-fed echelle spectrograph installed in 2003 at the 3.6\,m  telescope of La Silla Observatory, Chile \citep{Mayor2003}. The spectra used in this work were collected between April 2007 (BJD=2454194.9) and September 2017 (BJD=2458027.5) with an exposure time of 900 s. In the treatment of these data, we performed a separate analysis of the spectra taken before and after May 2015. This is because on that date the vacuum vessel that contains the spectrograph was opened to upgrade the fibre link \citep{LoCurto2015}, creating a discontinuity in the RV and index values, which necessitates of calculating an offset between the ``Pre-2015'' and ``Post-2015'' values. The instrument used for the wavelength calibration was a Thorium-Argon lamp \citep{Lovis2007}, which provides a large number of spectral lines distributed in the visible spectral range of HARPS. For the most recent data, we used an ultra-stable Fabry-Perot interferometer \citep{Wildi2010}, which provides the best short-term accuracy in radial velocity determination from the instrument.

HARPS-N is the northern counterpart of HARPS. This instrument was installed in 2012 at the 3.6\,m  Telescopio Nazionale Galileo (TNG) in Roque de los Muchachos Observatory, Spain \citep{Cosentino2012}. It has the same resolution as HARPS, similar wavelength coverage and is also contained in a vacuum vessel to minimize the temperature and pressure variations that may cause spectral drifts. The spectra used were taken between July 2014 (BJD=2456841.5) and October 2017 (BJD=2458038.4) with the same exposure time used in HARPS. The wavelength calibration was also done using a Th-Ar lamp.
 
CARMENES (Calar Alto high-Resolution search for M dwarfs with Exoearths with Near-infrared and optical Echelle Spectrographs) is a second generation echelle spectrograph installed in 2015 at the 3.5\,m telescope in Calar Alto Observatory, Spain \citep{Quirrenbach2018}. This instrument has two different channels that work simultaneously in the visible and near-infrared, and it is mainly focused on searching for low-mass planets in the habitable zones of late-type stars. The spectra used were acquired between February 2016 (BJD=2457422.7) and October 2017 (BJD=2458032.3), and we only use the visible channel. The calibration method is similar to the one used in HARPS, along with simultaneous Fabry-Perot exposures and a daily calibration using Th-Ne, U-Ar and U-Ne lamps \citep{Quirrenbach2018}.

UVES (Ultraviolet-Visual Echelle Spectrograph) is a high-resolution optical spectrograph installed in 2000 at the 8.2\,m VLT in Paranal Observatory, Chile \citep{Dekker2000}. The spectra used in this work were taken between April 2003 (BJD=2452743.4) and October 2005 (BJD=2453658.0). These spectra were acquired using an image slicer, with an effective slit width of 0.3 arcsec that gives a resolution of $\sim$130 000. UVES data are calibrated using the standard Th-Ar lamp, and in addition, accurate RVs are extracted thanks to an additional calibration based on an Iodine Cell, which provides many absorption lines on top of the target spectrum in some spectral regions. This makes some parts of the spectra not useful to measure for instance certain chromospheric indexes like the Na I D. 

HIRES (HIgh-Resolution Echelle Spectrometer) is a first generation echelle spectrograph installed in 1996 at the 10\,m Keck telescope in Mauna Kea Observatory, USA \citep{Vogt1994}. The spectra used were collected between August 2004 (BJD=2453237.9) and September 2014 (BJD=2456908.7). The wavelength calibration was done in a similar way as for the UVES spectra, i.e. inserting the Iodine Cell in the light beam with the aim of improving the RV precision.

PFS (Carnegie Planet Finder Spectrograph) is a high-resolution optical echelle spectrograph installed in 2009 at the 6.5\,m  Magellan II telescope in Las Campanas Observatory, Chile \citep{Crane2010}. The spectra we use were taken between August 2011 (BJD=2455791.6) and August 2016 (BJD=2457615.6). The wavelength calibration method is the same as the one used in UVES and HIRES.

The APF (Automated Planet Finder) consists of a 2.4\,m  telescope with the Levy Spectrometer commissioned in 2013 at the Lick Observatory, USA \citep{Vogt2014}. The spectra we use were acquired between July 2013 (BJD=2456504.7) and March 2016 (BJD=2457478.0). This instrument has a similar optical configuration to PFS, and also uses an Iodine Cell to make the wavelength calibration.

\subsection{Photometric Dataset}

For the photometric analysis, we relied on data taken with four different sources. Archival publicly available data come from the ASAS survey, which has a time base of several years:

ASAS (All Sky Automated Survey) consists of two automated observing stations at Las Campanas Observatory, Chile (ASAS-S or ASAS-3), and Haleakal\={a} Observatory, USA (ASAS-N or ASAS-3N) \citep{Pojmanski1997}. These two stations observe simultaneously in the \textit{V} and \textit{I} photometric bands with an average accuracy of $\sim$ 0.05 mag per exposure. They are complemented with the ASAS-SN (All-Sky Automated Survey for Supernovae) project \citep{Shappee2014}, which consists of 20 telescopes distributed around the globe that are automatically surveying the entire available sky every night down to \textit{V} $\sim$ 17 mag. Data from the ASAS-S and ASAS-SN were retrieved from its public database\footnote{\url{http://www.astrouw.edu.pl/asas/}}, while data from ASAS-N were supplied by M. Kiraga (priv. comm.), as they have not yet been made public. We thus collect 836 epochs (measurements averaged to one per night) from this survey (coming from ASAS-N, ASAS-S, and ASAS-SN) that were taken between September 2002 (BJD=2452524.6) and October 2017 (BJD=2458032.7).

Our own data comprise the second longest dataset of all, after
ASAS, coming from the Four College Automated Photoelectric Telescope (FCAPT) and the Robotically-Controlled Telescope (RCT), with a time span covering 14.5 years:

The FCAPT is a 0.75\,m automated telescope installed at the Fairborn Observatory (USA) that provides differential Str\"omgren uvby, Johnson \textit{BV}, and Cousins \textit{RI} photometry of a wide variety of stars \citep{Adelman2001}. RCT is a 1.3\,m telescope installed at the Kitt Peak National Observatory (USA) that includes a \textit{UBVRI} broadband filter set and is focused on observing faint objects such as brown dwarfs \citep{Gelderman2001}. The combined dataset from these two instruments is composed by 348 epochs, acquired in the V Johnson filter, which were taken between May 2003 (BJD=2452764.0) and June 2017 (BJD=2457922.8).

In addition, we orchestrated a joint photometric follow-up campaign for Barnard's Star, quasi-simultaneous to its Doppler observations acquired as part of the Red Dots 2017 (RD2017) campaign \footnote{\url{https://reddots.space/}}, designed to search for planet signatures around our closest M dwarf neighbors. The participating observatories were:

The Sierra Nevada Observatory (SNO, Spain), whose data come from the 0.9\,m telescope (T90) that is equipped with a CCD camera VersArray 2Kx2K with a 13.2$\times$13.2\,arcmin$^2$ field of view. We work with 69 epochs from this telescope that were taken since May 2017 (BJD=2457887.6) until October 2017 (BJD=2458042.3), quasi-simultaneous to the RD2017 campaign. We collected about 30 measurements per night in each in \textit{B}, \textit{V} and \textit{R} Johnson filters, accounting for a total of about 2000 observations in each filter. 

The Montsec Astronomical Observatory (OAdM, Spain), whose data come from the Joan Or\'o robotic telescope (TJO) that is equipped with a CCD Andor DW936N-BV camera with a 12.3$\times$12.3\,arcmin$^2$ field of view. We work with 72 epochs from this telescope that were taken since June 2017 (BJD=2457920.5) until October 2017 (BJD=2458050.3), quasi-simultaneous to the RD2017 campaign. A minimum of 5 measurements was done per night, to finally obtain a total of about 700 images in two filters (\textit{R} and \textit{I}). As the majority of photometric data from other instruments were acquired in the V filter, we do not include the OAdM dataset in the final analysis.

Following the outreach spirit of the Pale Red Dot campaign \citep{Anglada-Escude2016}, our desire was that the RD2017 campaign involved as many members of the public as possible. Thus, in addition to the setup of the RD2017 website \addtocounter{footnote}{-1}\addtocounter{Hfootnote}{-1}\footnotemark \ and social media for the campaign, we requested support from the AAVSO (American Association of Variable Stars Observers) and issued an AAVSO alert with a call for photometric follow-up from observers. The answer was enthusiastic, with more than 8000 measurements in the \textit{BVRI} and H$\alpha$ filters for Barnard's Star uploaded to the AAVSO database from 14 observers in eight countries (see Table \ref{tab:aavso}). About 75\% of the observations/acquired exposures (or half of the datasets) had great quality and could be included in the analysis, covering a time-span of 120 days with 6310 measurements in 148 epochs, as measurements from different observers were not consolidated into nightly binned epochs. 

We also analysed data from the Las Cumbres Observatory network (LCO) and the ASH2 0.40\,m telescope at SPACEOBS (San Pedro de Atacama Celestial Explorations Observatory) observatory (Chile), the latter being operated by the Instituto de Astrof\'isica de Andaluc\'ia (IAA).

In the case of LCO, data were obtained in the \textit{B} and \textit{V} Johnson and \textit{r'} and \textit{i'} Sloan filters. Unfortunately, data in the \textit{B}, \textit{r'} and \textit{i'} filters could not be used due to instrumental issues. The data in the \textit{V} filter were not included either in the final combined dataset due to their high dispersion both intra- and night-to-night, as reflected in the high mean error and root-mean-square (RMS), of 16.0\,mmag and 30.5\,mmag, respectively. The scattering was very high in comparison to the other observatories simultaneously acquiring data in the RD2017 campaign.

In the case of SPACEOBS, observations were acquired in three narrow-band filters with a FWHM of 12\,nm, centered on the OIII (501\,nm), SII (672\,nm) and H$\alpha$ (656\,nm) lines, with mean errors in the range of 14 to 24\,mmag, larger than in most datasets. This is most likely attributed to the narrow filters and faint comparison stars. The night-to-night stability, shown by the RMS, is low, with values ranging from 7 to 13\,mmag, depending on the filter. The narrow band lines were useful to monitor any possible activity bursts, such as flares, but were not included in the final combined dataset due to the short time base and larger scatter compared to the other RD2017 observatories simultaneously acquiring data.

Finally, we also analyse publicly available data from the MEarth survey, which consists of two arrays of robotically controlled telescopes located at the Fred Lawrence Whipple Observatory (USA) and Cerro Tololo Inter-American Observatory (Chile) \citep{Berta2012}. Each array consists of eight identical telescopes with a 0.4\,m primary mirror that focuses the light onto a high-grade CCD camera with a broad RG715 nm filter. We work with 161 epochs from this survey that were taken since February 2013 (BJD=2454876.0) until October 2015 (BJD=2457323.6). The large mean error in this dataset indicates that the measurements had a large intra-night scatter, but once consolidated into nightly averages, the scatter of the whole run decreased to 6.5\,mmag, indicating that there were not large differences from night to night observations. We did not combine this dataset with the rest of the time-series because it was taken with a filter that did not match the \textit{V} Johnson filter used in the other datasets.

The properties of all of these photometric sources are shown in Table \ref{tab:Photometric_Instruments}, including the mean error of the averaged nights, which indicates the scatter of the measurements within the night, giving an idea of the quality of the nights; and the RMS of the run, which gives a measure of the night-to-night stability. We mark in boldface the selected datasets that we used for the analysis presented in this paper (using the \textit{V}-filter time-series in each case).

\begin{table}
\centering
	\caption{Properties of the photometric data.}
	\label{tab:Photometric_Instruments}
	\begin{tabular}{lcccc} 
		\hline
		Observatory/      & \multirow{2}{*}{Aperture} & \multirow{3}{*}{Filter}    & \multirow{2}{*}{Error} & \multirow{2}{*}{RMS}    \\
		Survey/           &  \multirow{2}{*}{[m]}     &                            & \multirow{2}{*}{[mmag]}   & \multirow{2}{*}{[mmag]}\\
		Telescope         &                           &                            &                           &                        \\
		\hline
		ASAS-3            & 0.07                      & \textit{V}                         & 10.3                      & 17.0 \\
		ASAS-3N           & 0.10                      & \textit{V}                         & 13.0                      & 16.1 \\
		ASAS-SN           & 0.14                      & \textit{V}                         & 5.2                       & 8.3  \\
        \textbf{Combined} & \multirow{2}{*}{0.07, 0.10, 0.14} & \multirow{2}{*}{\textit{V}}      & \multirow{2}{*}{10.4}     & \multirow{2}{*}{16.8}  \\
        \textbf{ASAS}     &                           &                                    &                           &                        \\
		\hline
        \textbf{FCAPT \&} & \multirow{2}{*}{0.80, 1.30} & \multirow{2}{*}{\textit{V}}      & \multirow{2}{*}{4.9}      & \multirow{2}{*}{11.2}  \\	
        \textbf{RCT}      &                             &                                  &                           &                        \\	
		\hline
		MEarth            & 0.40                      & \textit{RG715}                     & 16.5                      & 6.5                    \\
		\hline
		\textbf{SNO}      & 0.90                      & \textit{B}                         & 4.5                       & 5.4                    \\
		                  &                           & \textit{V}                         & 4.4                       & 6.4                    \\
		                  &                           & \textit{R}                         & 5.8                       & 5.3                    \\
		\hline
		OAdM              & 0.80                      & \textit{R}                         & 7.2                       & 9.6                    \\
		                  &                           & \textit{I}                         & 8.4                       & 8.8                    \\	
		\hline		           	
		\textbf{AAVSO}    & Range                     & \textit{V} & 15.1                      & 8.9                    \\
	    \hline		
	    LCO               & 0.40                      & \textit{V}                          & 16.0                      & 30.5                   \\
	                      &                           & \textit{r'}                   & 31.1                      & 45.2                   \\
	                      &                           & \textit{i'}                  & 91.4                      & 75.6                   \\
 	    \hline		           
	    ASH2              & 0.40                      & [OIII]                       & 14.1                      & 7.1                    \\
	                      &                           & H$\alpha$                 & 23.9                      & 12.5                   \\	
	                      &                           & [SII]                        & 16.8                      & 9.5                    \\	
		\hline
	\end{tabular}
\begin{minipage}{\columnwidth} 
{\footnotesize \textbf{Columns:} Observatory, survey or telescope; telescope aperture; filters; mean error of the nights and RMS of the run (see the text for details).}
\end{minipage}	
\end{table}

All photometric data (except for ASAS and MEarth) were reduced with standard procedures including bias and/or dark subtraction and flat-field correction. Several apertures were tried to extract the best aperture photometry that maximized the signal-to-noise ratio (SNR). Differential magnitudes were obtained with respect to nearby comparison stars that had previously been checked for stability and, in the case of observations taking place during the RD2017 campaign, agreed upon, so that the different photometric datasets were as uniform and comparable as possible.

\section{Method}

\label{sec:Method}

\subsection{Determination of Stellar Activity Indicators}

In order to measure activity indices, we first correct all the spectra for the blaze function. In the case of HARPS and HARPS-N we use a specific blaze spectrum given by their respective pipelines, and for the other spectrographs we fit a second order polynomial to each order to create an artificial blaze spectrum.

Next, we correct for the pixel size variability in wavelength, which requires to re-binning the spectra to obtain a constant step in wavelength between pixels and also to correct accordingly the flux evaluated in the selected wavelength step (0.01 \AA). Then we correct the wavelength for the barycentric velocity of the Earth and the radial velocity of the star. Both velocities are available in the header of the HARPS, HARPS-N, and CARMENES spectra. In the case of HIRES, APF, PFS, and UVES, we calculate the barycentric velocity using the equatorial coordinates (RA and Dec) and the Julian day (BJD), and we use a calculated value of -110.25 km/s for the radial velocity (obtained by averaging the HARPS and HARPS-N header values). To deal with the small wavelength shifts (few m/s) related to the use of a mean value for the radial velocity instead of variable value over time, we correlate the spectra in these four spectrographs using the first spectrum of each spectrograph as reference. Finally, we build an average spectrum and use the individual spectra to calculate the weights of each echelle order involved in the index. The weight of one order in a certain spectrum is calculated as the quotient between the normalized median of the flux of this order in the average spectrum and the selected spectrum.

\begin{figure} 
	\includegraphics[width=\columnwidth]{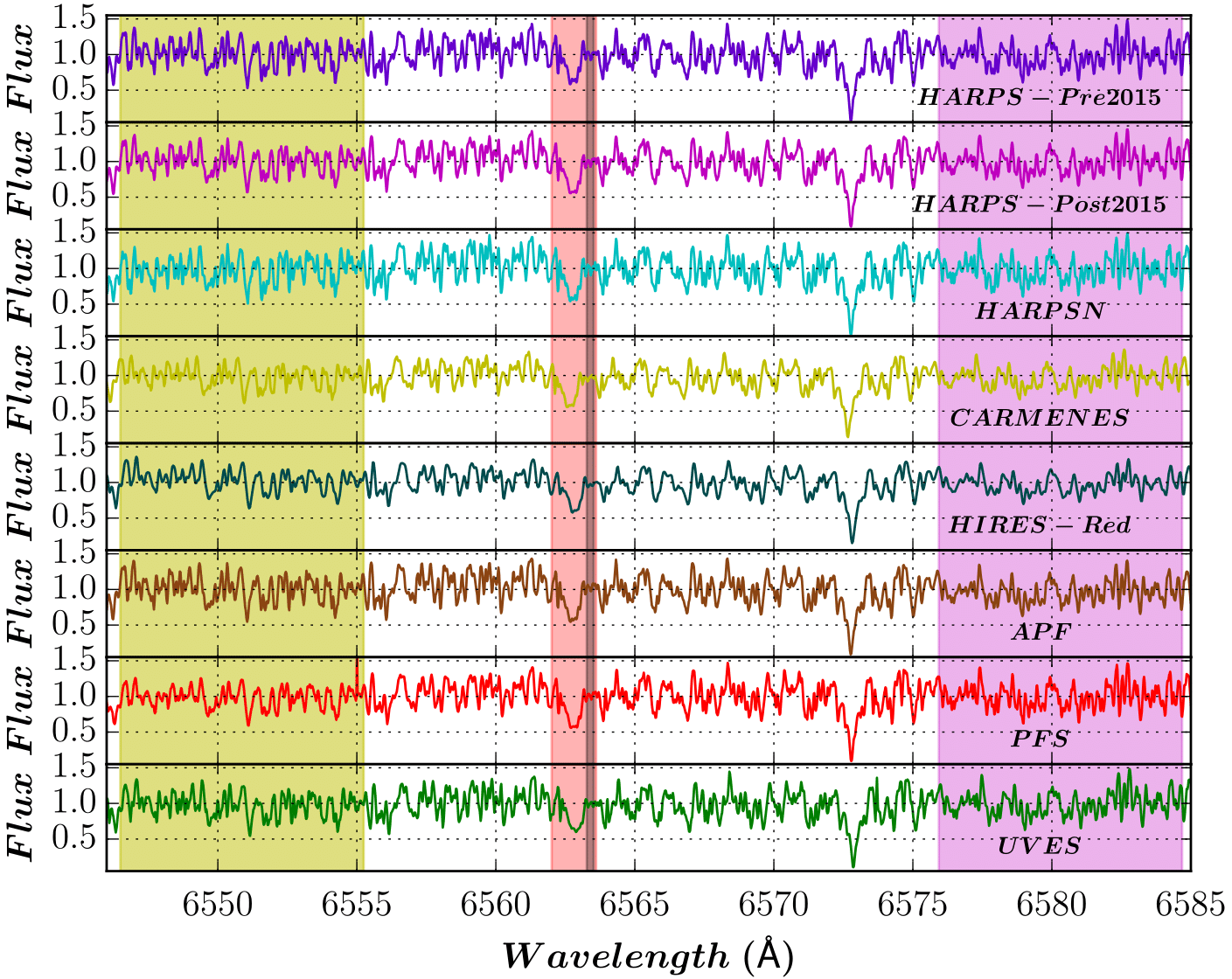} 
    \caption{Normalized one-dimensional spectra taken with seven spectrographs. The H$\alpha$ band is marked in pink, the continuum passbands are marked in yellow and violet and the continuum region used to calculate the index error is marked in grey.}
    \label{fig:Halpha_AllSpectrographs}
\end{figure}

Once all the spectra have passed through this process, we can measure the three activity indices. The first one is the H$\alpha$ index, which we define as:

\begin{equation}
  H\alpha=\frac{A}{L+R}
  \label{eq:Halpha_index}
\end{equation}

\noindent where \textit{A} is a rectangular passband centered at the core of the H$\alpha$ line (6562.808 \AA) with a width of 1.6 \AA, and \textit{L} and \textit{R} are the continuum bands centered at 6550.870 and 6580.310 \AA \ respectively, with a width of 8.75 \AA \ \citep{Kurster2003,Gomes2011}. Fig. \ref{fig:Halpha_AllSpectrographs} shows this spectral region for the seven spectrographs in which it is possible to measure this index.

\begin{figure} 
	\includegraphics[width=\columnwidth]{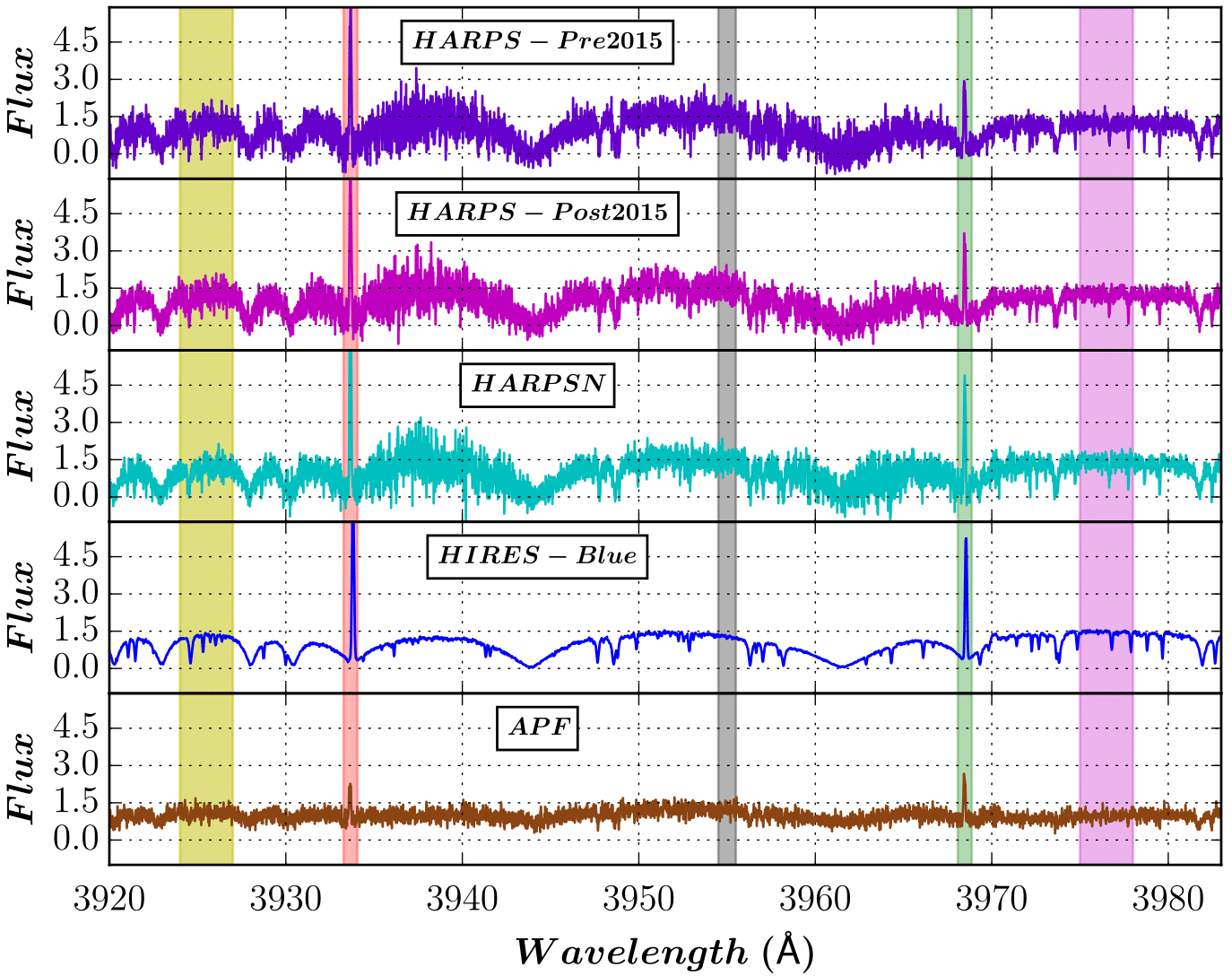} 
    \caption{Normalized one-dimensional spectra taken with four spectrographs. The Ca II H\&K bands are marked in green and pink respectively, the continuum passbands are marked in yellow and violet and the continuum region used for the index error is marked in grey.}
    \label{fig:Smw_AllSpectrographs}
\end{figure}

The second one is similar to the S-index related to the CaII H \& K lines \citep{Noyes1984}, that we call CaHK index and define as:

\begin{equation}
  S=\frac{H+K}{R+V}
  \label{eq:S_index}
\end{equation}

\noindent where \textit{H} and \textit{K} are triangular passbands for the core of the lines (centered at 3968.470 and 3933.664 \AA, respectively) with a full width at half maximum (FWHM) of 1.09 \AA. In this work we have shifted the continuum filters of \textit{R} and \textit{V} from 4001.070 and 3901.070\,\AA, to 3976.5 and 3925.5 \AA, respectively, and also modified the width of both filters from 20 to 3\,\AA, in order to use narrower spectral regions near the core of the lines located in the same echelle orders as those lines. These continuum bands allow us to avoid the overlap between different echelle orders in all of the spectrographs. Fig. \ref{fig:Smw_AllSpectrographs} shows this spectral region for the four spectrographs in which it is possible to measure this index.

\begin{figure} 
	\includegraphics[width=\columnwidth]{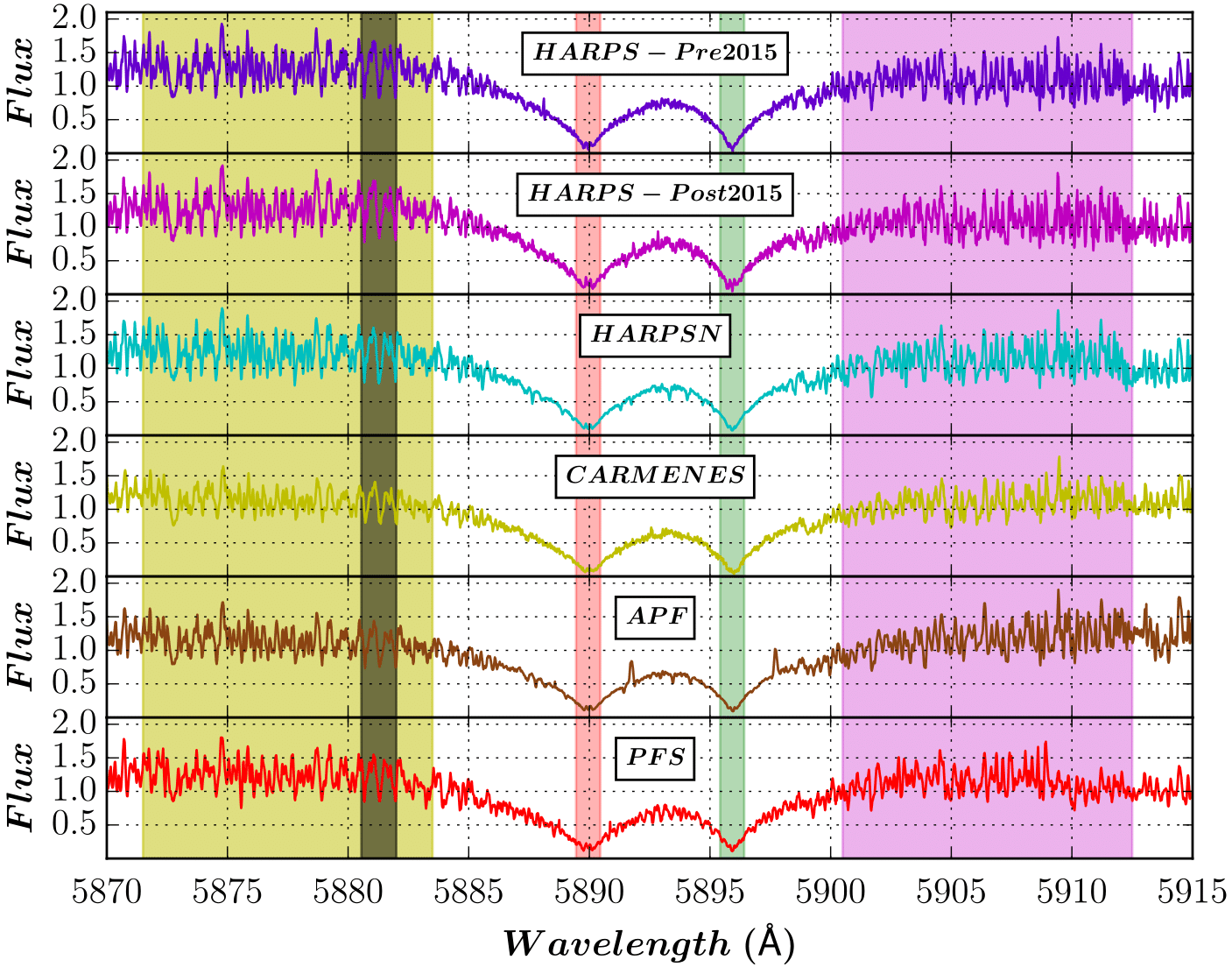} 
    \caption{Normalized one-dimensional spectra taken with five spectrographs. The Na I D$_{1}$ and D$_{2}$ bands are marked in pink and green respectively, the continuum passbands are marked in yellow and violet and the continuum region used for the index error is marked in grey.}
    \label{fig:Na_AllSpectrographs}
\end{figure}

The last activity indicator is the Na I D index \citep{Díaz2007}, which we define as:

\begin{equation}
  N=\frac{D_{1}+D_{2}}{L+R}
  \label{eq:Na_index}
\end{equation}

\noindent where \textit{D$_{1}$} and \textit{D$_{2}$} are rectangular passbands for the core of the sodium doublet lines (centered at 5895.92 and 5889.95 \AA, respectively) with a width of 1 \AA. \textit{L} and \textit{R} are the continuum bands that are usually centered at 5805.0 and 6090.0 \AA, with a width of 10 and 20 \AA, respectively, but in this work we shifted them to 5881.5 and 5902.5 \AA. We also modified their widths to 12 \AA \ for the same reason as the one used for the CaHK index. Fig. \ref{fig:Na_AllSpectrographs} shows this spectral region for the five spectrographs in which it is possible to measure this index.

The uncertainties of the three indices were determined through error propagation \citep{Taylor1982}, using the RMS of the error region marked in grey in Fig. \ref{fig:Halpha_AllSpectrographs}, \ref{fig:Smw_AllSpectrographs} and \ref{fig:Na_AllSpectrographs} as the error for the bands $A$, $L$, $R$, $H$, $K$, $V$, $D_{1}$ and $D_{2}$.
  
\begin{figure} 
	\includegraphics[width=\columnwidth]{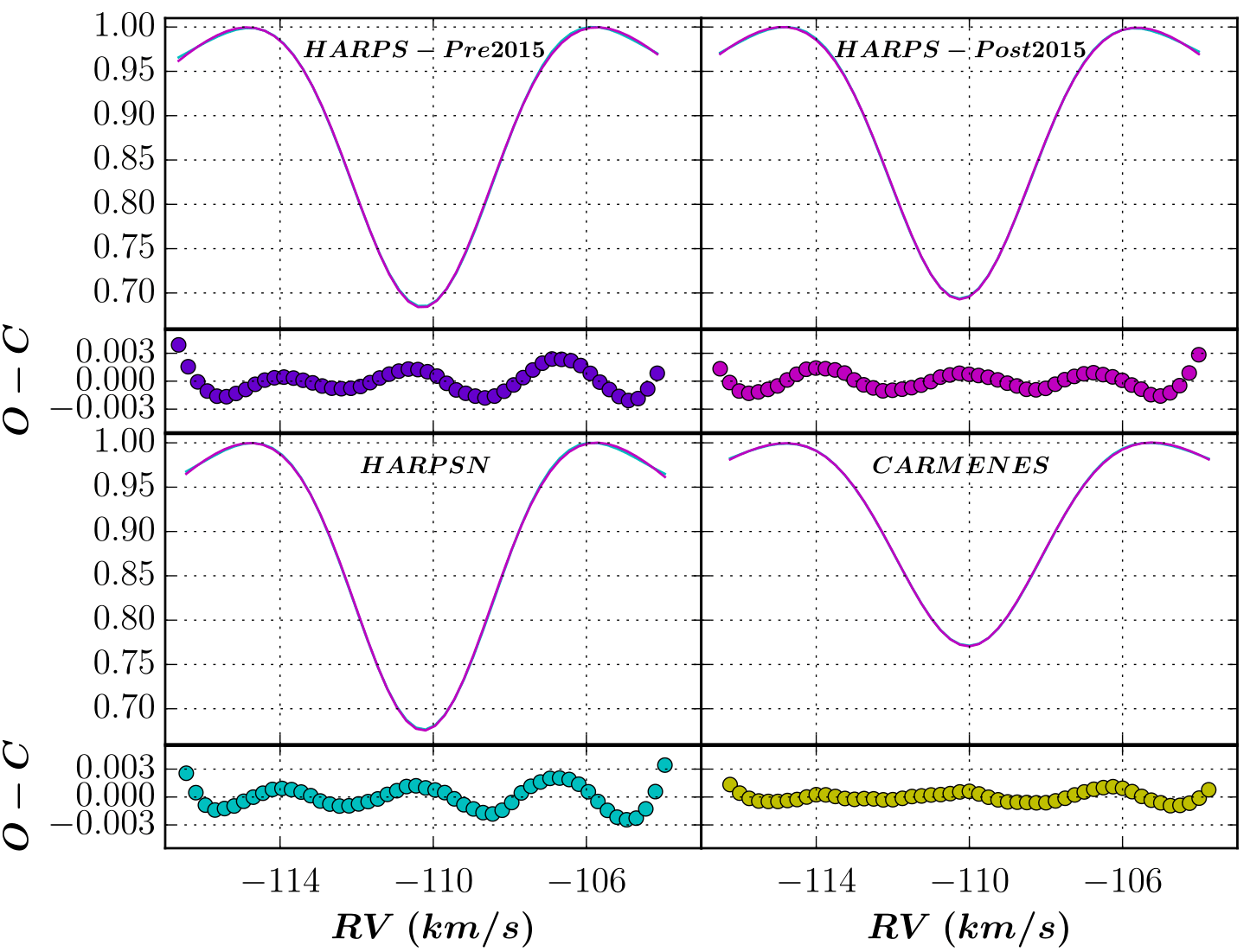} 
    \caption{CCFs obtained for a single observation of three different spectrographs along with their respective residuals.}
    \label{fig:FWHM_AllSpectrographs}
\end{figure}

We also used the cross-correlation function (CCF) computed by the HARPS, HARPS-N, and CARMENES pipelines to estimate the FWHM as an additional activity indicator. We computed an average CCF using individual weights for each echelle order as we did with the spectra (we build an average CCF and use the individual CCFs to calculate the weights of each echelle order). Then we cut the CCF to a width of 25 pixels and used a Gaussian+second-order-polynomial fit to obtain the FWHM. In Fig. \ref{fig:FWHM_AllSpectrographs} four CCF fits are shown, one per spectrograph, along with their residuals.

Once we had the measurements from all the spectra, we computed the weighted average per night, discarding those values that are beyond 3$\sigma$ from the median index in order to remove outliers. We also discarded values with errors beyond 3$\sigma$ from the median error. The outliers may be associated in some cases with flares, a phenomenon already detected in Barnard's Star \citep{Paulson2006}, though occurring rarely due to its advanced age. We also applied this treatment to the photometric data, in which we already had a set of magnitudes measured with different instruments. This process gave the final number of datapoints shown in Table \ref{tab:Nspec} for the four spectroscopic indices and the photometric magnitudes.

\begin{table}
\centering
	\caption{Number of measurements used for every index after the selection criteria were applied.}
	\label{tab:Nspec}
	\begin{tabular}{lccccc}
		\hline
		\multirow{2}{*}{Instrument} & \multicolumn{4}{c}{N$_{\rm measurements}$} \\
		& H$\alpha$ & CaHK & NaD & FWHM & $m_{\rm V}$ \\
		
		\hline
		HARPS-Pre2015  & 109 & 110 & 114 & 115 & ... \\
		HARPS-Post2015 & 66  & 66  & 63  & 65  & ...  \\
		HARPS-N        & 40  & 39  & 40  & 40  & ...  \\
		CARMENES       & 182 & ... & 164 & 173 & ... \\
		HIRES          & 124 & 125 & ... & ... & ... \\
		APF            & 44  & 45  & 42  & ... & ... \\
		PFS            & 33  & ... & 30  & ... & ... \\
		UVES           & 21  & ... & ... & ... & ... \\
	    ASAS           & ... & ... & ... & ... & 830 \\
	    FCAPT-RCT      & ... & ... & ... & ... & 344 \\
	    AAVSO          & ... & ... & ... & ... & 148 \\
	    SNO            & ... & ... & ... & ... & 68 \\	
		\hline
		Combined       & 619 & 385 & 453 & 393 & 1390 \\	
		\hline
	\end{tabular}
\end{table}

The relative offsets between instruments were calculated for each index separately. We divided the spectroscopic data into two separate blocks according to their time-span in order to have enough overlapping observations: the first one included HIRES, HARPS-Pre2015, PFS, APF, and UVES; and the second one included HARPS-Post2015, HARPS-N, and CARMENES. We used time windows of 10 days for spectrographs of the same block, and 30 days for spectrographs of different blocks. We determined the difference between the values contained in these windows and averaged all of them to obtain the offset. For the photometric data, we only used one block of instruments due to the long-time coverage of surveys like ASAS. These offsets are shown in Table \ref{tab:Offsets1}.

\begin{table*}
\centering
	\caption{Offsets between spectral and photometric indices datasets from different instruments.}
	\label{tab:Offsets1}
	\begin{tabular}{lccccc}
		\hline
		Instruments & H$\alpha$ Offset & CaHK Offset & NaD Offset & FWHM Offset [km/s] & $m_{\rm V}$ Offset [mag] \\
		\hline
		HIRES-HARPSpre     & 0.00198 $\pm$ 0.00006  & 0.4817 $\pm$ 0.0006  & ...                    & ...                   & ... \\
		HIRES-PFS          & 0.01357 $\pm$ 0.00008  & ...                  & ...                    & ...                   & ... \\
		HIRES-APF          & -0.024  $\pm$ 0.002    & -2.276 $\pm$ 0.007   & ...                    & ...                   & ... \\
		HIRES-UVES         & 0.05624 $\pm$ 0.00006  & ...                  & ...                    & ...                   & ... \\
		CARMENES-HARPSpost & -0.01547 $\pm$ 0.00002 & ...                  & -0.00678 $\pm$ 0.00007 & 0.12888 $\pm$ 0.00001 & ... \\
		CARMENES-HARPSN    & -0.03559 $\pm$ 0.00003 & ...                  & -0.00412 $\pm$ 0.00005 & 0.20562 $\pm$ 0.00001 & ... \\
		HARPSpost-HARPSN   & ...                    & -0.5138 $\pm$ 0.0006 & ...                    & ...                   & ... \\
		HARPSpre-PFS       & ...                    & ...                  & 0.0250 $\pm$ 0.0002    & ...                   & ... \\
		HARPSpre-APF       & ...                    & ...                  & -0.075 $\pm$ 0.002     & ...                   & ... \\
		ASAS-FCAPT+RCT     & ...                    & ...                  & ...                    & ...      & -0.00870 $\pm$ 0.00002 \\
		ASAS-AAVSO         & ...                    & ...                  & ...                    & ...      & -0.01531 $\pm$ 0.00002 \\
		ASAS-SNO           & ...                    & ...                & ...                    & ...      & -0.011340 $\pm$ 0.000007 \\
		\hline
	\end{tabular}
\end{table*}

\begin{figure*} 
	\includegraphics[width=17.5cm]{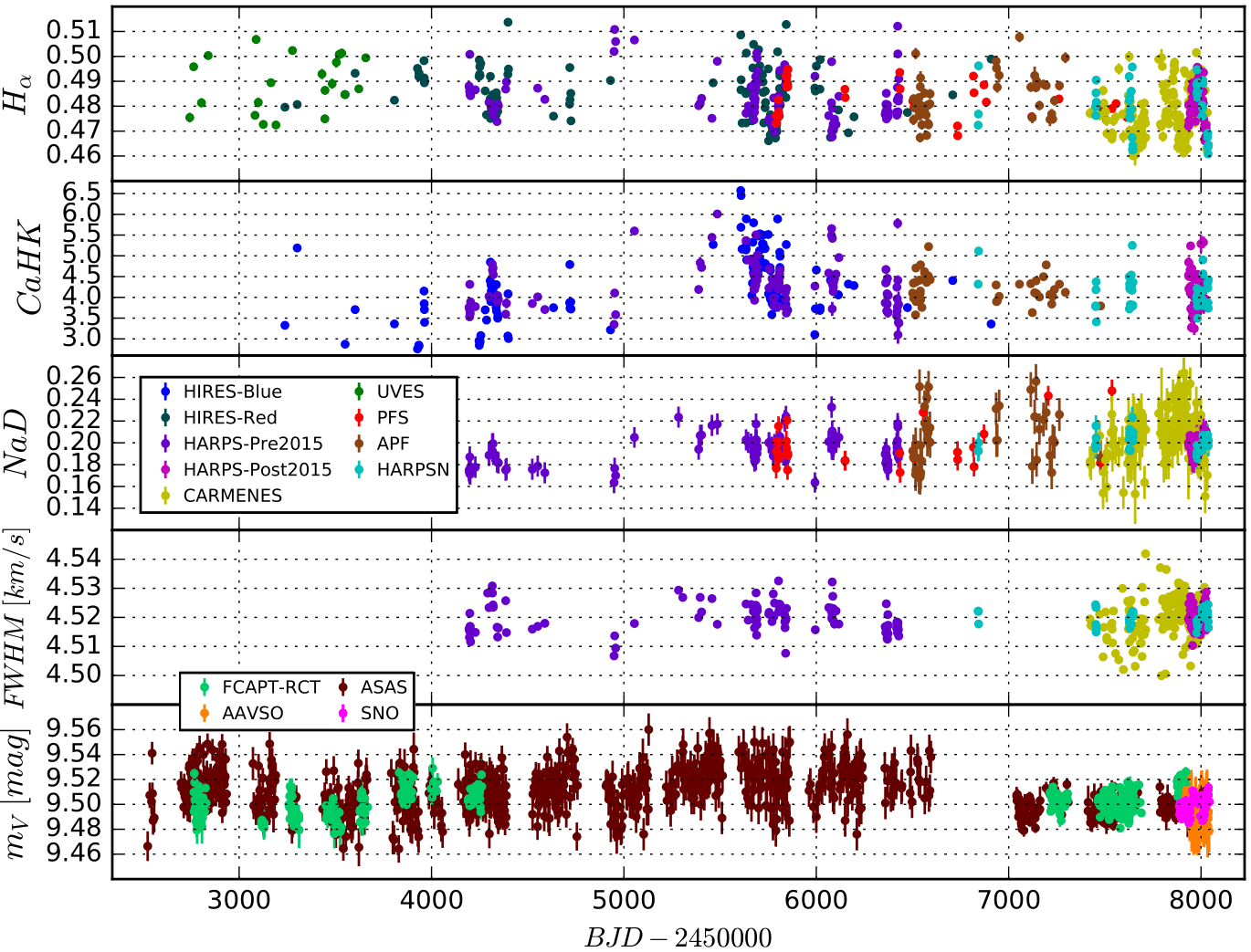} 
    \caption{Time series of the four spectroscopic indexes and the \textit{V} photometry with their respective offsets applied. The NaD plot contains a legend with all the spectrographs and the FWHM plot contains a legend with the instruments used for the photometry analysis.}
    \label{fig:Compact_Halpha_Smw_Na}
\end{figure*}

After applying these offsets, we apply another 3$\sigma$-clipping to the complete dataset values and we get the time-series (of each spectroscopic index and photometric magnitude) shown in Fig. \ref{fig:Compact_Halpha_Smw_Na}.

\subsection{Time Series Analysis}

We carry out a time-series analysis of the three spectroscopic activity indicators, the CCF FWHM and the \textit{V}-band photometry using the Lomb-Scargle periodogram \citep{Lomb1976} in its generalized form \citep{Zechmeister2009}, in which each value has an independent error. We also analyse the chromatic index given only in the CARMENES data. Each point of the periodogram is calculated as:

\begin{equation}
  z(\omega)=\frac{N-3}{2} \cdot p(\omega)=\frac{N-3}{2} \cdot \frac{\chi_{o}^{2}-\chi^{2}(\omega)}{\chi_{o}^{2}}
  \label{eq:zw}
\end{equation}

\noindent where $N$ are the degrees of freedom and $\chi^{2}$ is the squared difference between the data and the model for a certain frequency $\omega$, calculated as:

\begin{equation}
  \chi^{2}=\sum_{i=1}^{N} \ \frac{[y_{i}-y(t_{i})]^{2}}{\sigma_{i}^{2}}
  \label{eq:chi}
\end{equation}

The False Alarm Probability (FAP) \citep{Horne1986} associated with every point in the periodogram is calculated with the following expression \citep{Cumming2004}: 

\begin{equation}
  FAP=1-[1-P(z>z_{o})]^{M}=1-[1-e^{-z_{o}}]^{M}
  \label{eq:FAP}
\end{equation}

\noindent where $z$ is the real power of a point in the periodogram,  $z_{o}$ is the measured power, $M$ is the number of independent frequencies used in the periodogram, and $P(z>z_{o})$ measures the probability of $z$ being greater than $z_{o}$. Using Eq.~(\ref{eq:FAP}) we established a first approximation of the 10\%, 1\% and 0.1\% levels of FAP for the periodograms. To obtain more precise values of these levels we applied a bootstrapping method \citep{Endl2001}. This method involves re-arranging the time order of the indices values 10000 times, searching for the period with the highest significance in every iteration to determine which values are obtained 10\%, 1\% and 0.1\% of the times.

\begin{figure} 
	\includegraphics[width=\columnwidth]{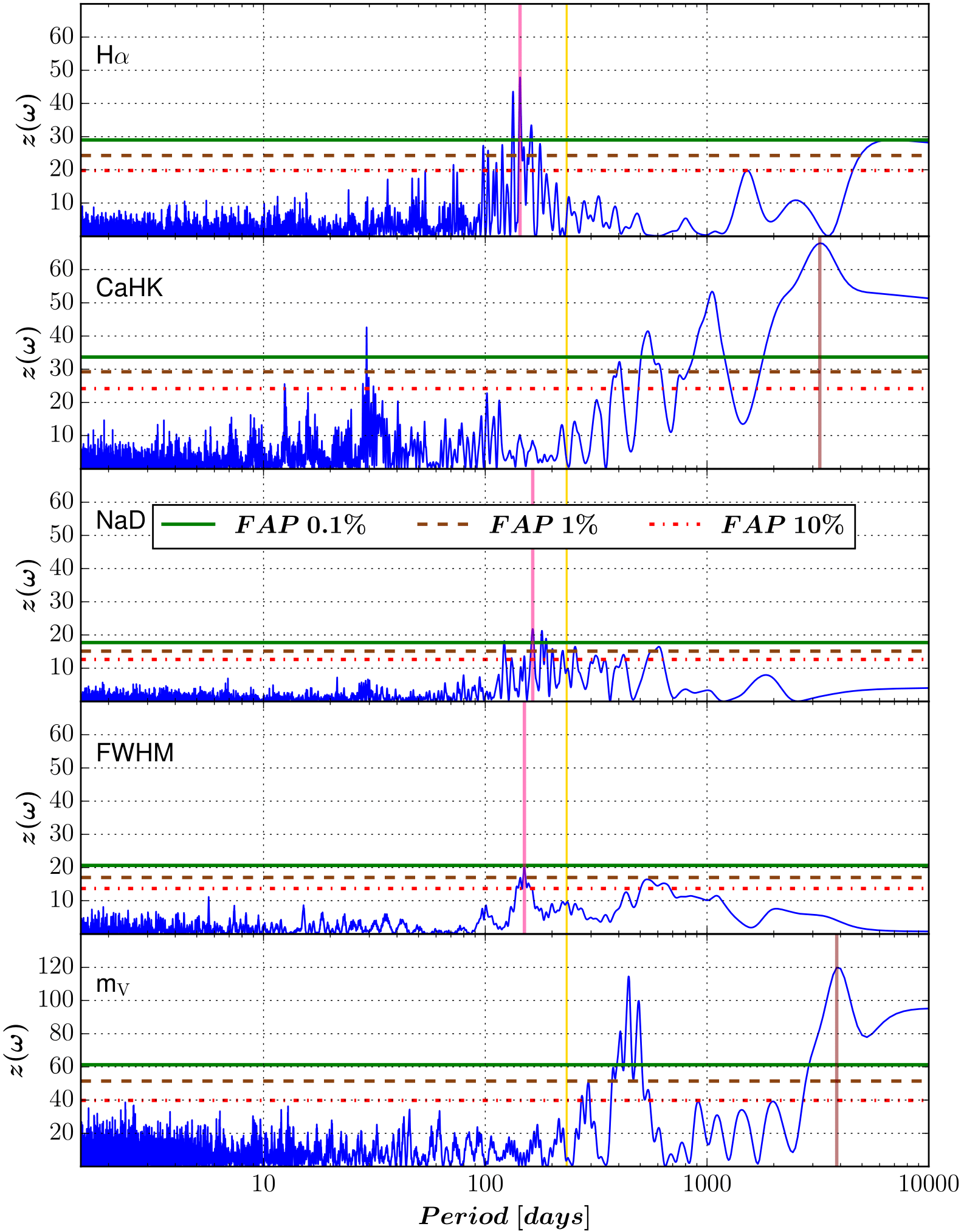} 
    \caption{Periodograms of the four spectroscopic indexes and the V-band photometry using the values from Fig. \ref{fig:Compact_Halpha_Smw_Na}. The most significant signals have been marked in different colors: pink for the ones associated with the rotation period and brown for the ones associated with the long-term activity cycle. The vertical yellow line shows the period of the recently discovered super-Earth Barnard b \citep{Ribas2018}.}
    \label{fig:Compact_Halpha_Smw_Na_FirstPeriodograms}
\end{figure}

In Fig. \ref{fig:Compact_Halpha_Smw_Na_FirstPeriodograms} we show the periodograms for the four spectroscopic indices and the \textit{V} magnitude using the time-series from Fig.\,\ref{fig:Compact_Halpha_Smw_Na} (which includes all the instruments used with their respective offsets, except in the case of the photometric time-series, where we did not use the FCAPT-RCT dataset for reasons that will be discussed later) with the FAP levels from bootstrapping. Signals with a FAP lower than 0.1\% (i.e. with a $z>z_{FAP=0.1\%}$) are statistically significant, while for those with a FAP between 0.1\% and 10\% we can not
ensure that they are not false positives, and therefore we may discuss some signals, in particular, those below or close to a FAP of 1\% as tentative signals.

We first carried out a pre-whitening process for a single instrument. We started calculating the periodogram with the FAP levels from bootstrapping. We selected a signal from the periodogram (usually the most significant one) and modelled it with a double sinusoidal fit to subtract that signal and, thus, recompute the periodogram \citep{Boisse2011}. The double sinusoidal fit is defined as:

\begin{equation}
  y(t)=A_{1} \cdot sin (\omega_{1}+\phi_{1})+A_{2} \cdot sin (\omega_{2}+\phi_{2})+A_{3}
  \label{eq:Double_sinusoidal}
\end{equation}

\noindent where $\omega_{2}=2\omega_{1}=2 \pi f/P$. We left $A_{1},A_{2},A_{3},\phi_{1},\phi_{2}$ and $P$ as free parameters, restricting the value of $P$ in a 15\% from the original period in the periodogram. We used this double sinusoidal model in order to account the asymmetry of some signals \citep{Berdyugina2005} with the MPFIT routine \citep{Markwardt2009}. We also add a jitter term associated with every individual instrument present in the complete dataset to this model in order to account possible bad estimations in the index errors, along with a trend correction if we detect long-term variations above our time coverage. After subtracting the first signal, we repeated the process (maintaining the same FAP levels) until we had no more significant signals in the periodogram. 

After the pre-whitening process we isolated each individual signal from the rest of signals that we subtracted along this process. We selected one signal at a time and used the frequencies from the rest to make a model. The subtraction of this model gave us an isolated periodogram, where we can check that the original period was not caused by effects of the other signals. We obtain this isolated periodogram for every single signal that was detected along the pre-whitening process.

Once we carried out the whole process for one single instrument, we added a second instrument with its respective offset and repeated the modeling-subtraction-isolation method, because the information provided by a single instrument may not be enough in terms of time-span or sampling. The addition of instruments follows the order shown in Table \ref{tab:OrderSpectrographs}: we analysed each block of instruments separately and then join them. When we combined these two blocks of instruments, we needed to estimate an additional offset using a wider time window (30 days). These offsets are also shown in Table \ref{tab:OrderSpectrographs}.

Finally, we computed the window function for each time-series of each activity indicator including the photometric and RV data using the Systemic console \citep{Meschiari2009}. We find only a few signals related to the daily sampling and the yearly periodicity of the observations (the most significant at 365 and 1850 days).

\begin{table*}
\centering
	\caption{Addition order of spectrographs for each individual index and offsets between spectral indices datasets from different blocks of spectrographs.}
	\label{tab:OrderSpectrographs}
	\begin{tabular}{lccc}
		\hline
		Index & Block & Spectrograph & Offset \\
		\hline
		\multirow{2}{*}{H$\alpha$} & 1 & HIRES-HARPSpre-PFS-APF-UVES & \multirow{2}{*}{0.0145 $\pm$ 0.0002} \\
		 & 2 & CARMENES-HARPSpost-HARPSN \\
		\hline
		\multirow{2}{*}{CaHK} & 1 & HIRES-HARPSpre-APF & \multirow{2}{*}{0.62 $\pm$ 0.02} \\
		 & 2 & HARPSpost-HARPSN  \\
		\hline
		\multirow{2}{*}{NaD} & 1 & HARPSpre-PFS-APF & \multirow{2}{*}{0.0226 $\pm$ 0.0006} \\
		 & 2 & CARMENES-HARPSpost-HARPSN  \\
		\hline
		\multirow{2}{*}{FWHM} & 1 & HARPSpre & \multirow{2}{*}{-0.003114 $\pm$ 0.000001} \\
		 & 2 & CARMENES-HARPSpost-HARPSN  \\
		\hline
	\end{tabular}
\end{table*}

\section{Analysis}

\label{sec:Analysis}

In this section we describe the analysis and results of each activity indicator. In the spectroscopic analysis, we have used all datasets available according to Table \ref{tab:Nspec}. In the photometric analysis we use again all datasets given in Table \ref{tab:Nspec} (see more details in Section 4.5). All the results shown in this section has been done following the methodology described in Section \ref{sec:Method}.

\subsection{H$\alpha$ index}

\begin{figure} 
	\includegraphics[width=\columnwidth]{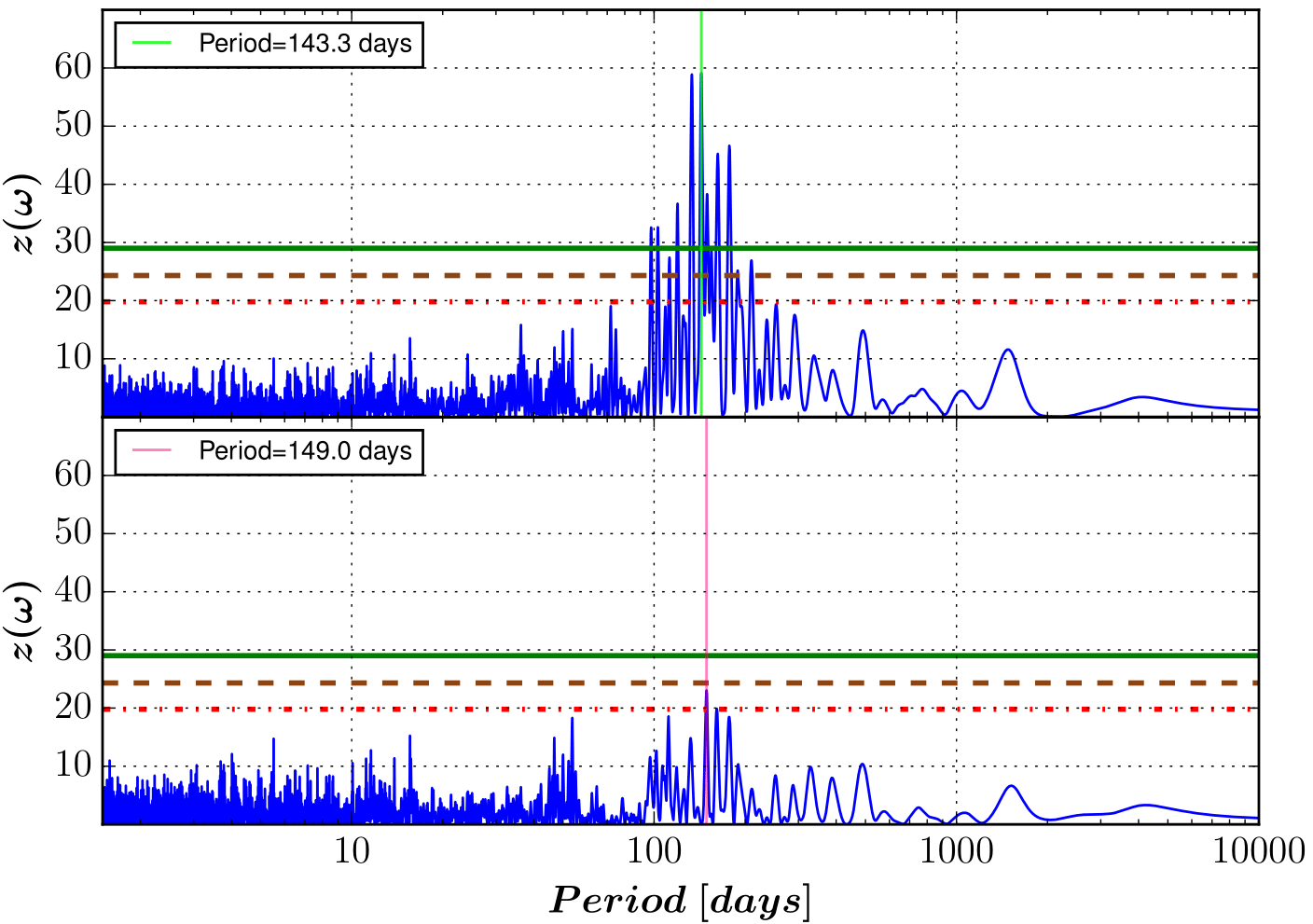}
    \caption{\textbf{Top:} Periodograms of the time-series of H$\alpha$ for HARPS+HARPSN+CARMENES+HIRES+APF+PFS+UVES spectra after the trend correction. \textbf{Bottom:} Periodogram of the residuals after the subtraction of the 143 days period signal.}
    \label{fig:Detect_allSignals_Halpha}
\end{figure}

In the case of H$\alpha$ we obtained 619 measurements, characterized by an average of 0.48, mean error of 0.001, and RMS of 0.01. We began analyzing the HIRES dataset (in the HIRES-Red configuration), and then added individually HARPS-Pre2015, PFS, APF, and UVES. After every addition, we used the modeling-subtraction technique to see which signals are hidden behind the main ones. We repeated this treatment for the second block of instruments, beginning with HARPS-Pre2015 and adding HARPS-N and CARMENES in that particular order. When we combined the two blocks, the periodogram of the data gives a 7692 days signal as the second most significant peak after the $\sim$ 140-150 days peak (see the first periodogram of Fig. \ref{fig:Compact_Halpha_Smw_Na_FirstPeriodograms}). To see if this signal was caused by any instrumental effect, we applied a trend correction to the whole dataset, and this signal disappeared, as it is shown in the top of Fig. \ref{fig:Detect_allSignals_Halpha}.

\begin{figure} 
	\includegraphics[width=\columnwidth]{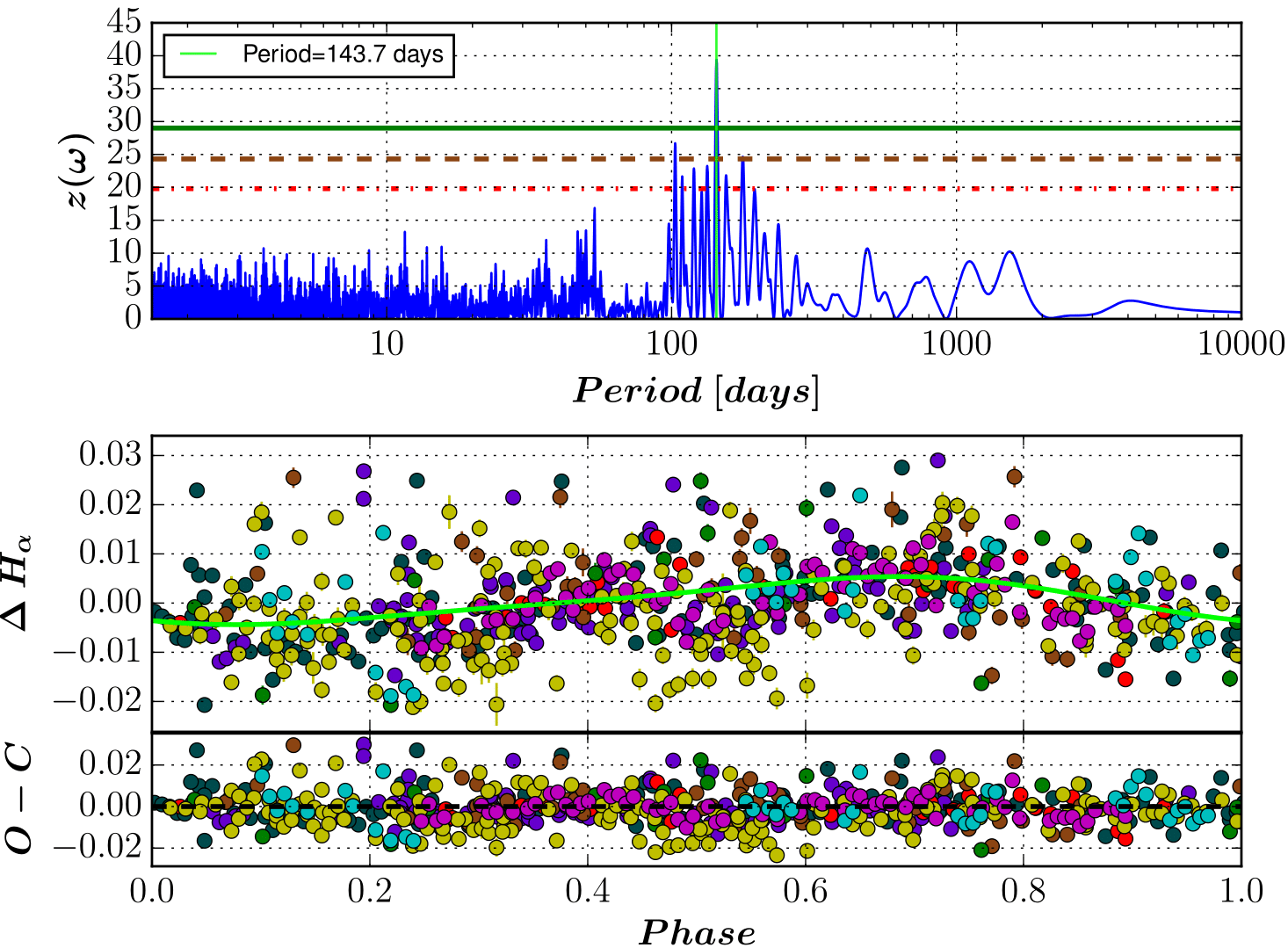}
    \caption{\textbf{Top:} Periodogram of the time-series of H$\alpha$ index after the subtraction of the 149 days period signal. \textbf{Bottom:} Phase-folded curve of the H$\alpha$ time-series using the 143 days period. Each spectrograph has been represented with a different color, following the legend in Fig. \ref{fig:Compact_Halpha_Smw_Na}. The green line represents the best double-sinusoidal fit found by the MPFIT routine.} 
    \label{fig:Periodogram_and_Fit_Halpha}
\end{figure}

After the trend correction, the most significant peak is at 143 days, which is close to the rotation period determined by \citet{Mascareño2015}. This signal is surrounded by multiple peaks between 130 and 177 days with low FAP. We fit this forest of peaks with a Gaussian model, whose FWHM gives us an error associated with the 143-day signal of 15 days. We note that the baseline of the observations is much longer than the expected lifetime of spots and plages on the surface of the star. These magnetic phenomena can occur at different stellar latitudes, favoring these multiple peaks around the rotation signal (see Section \ref{sec:Discussion}). In the second periodogram of Fig. \ref{fig:Detect_allSignals_Halpha}, the second signal detected, after the subtraction of the 143 days signal (modeled by a double sinusoidal), has a 149 days period with a FAP close to the 1\% level. When we isolate the first signal from the second one, the highest peak stays at 143 days with an amplitude of 0.00523\,$\pm$\,0.00001 and a FAP above the 0.1\%, as shown in Fig. \ref{fig:Periodogram_and_Fit_Halpha}.

\begin{figure} 
	\includegraphics[width=\columnwidth]{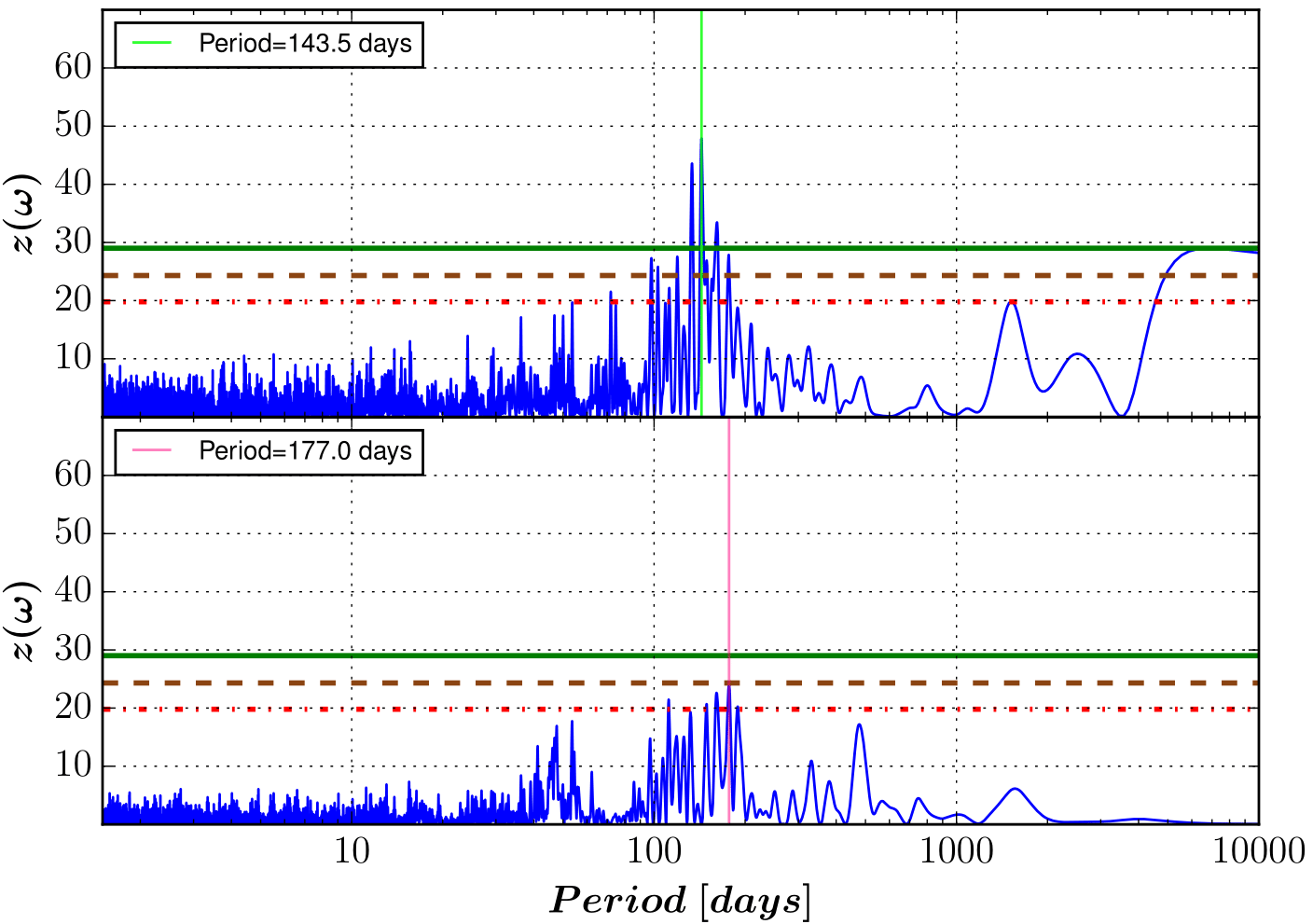}
    \caption{\textbf{Top:} Periodograms of the time-series of H$\alpha$ for HARPS+HARPSN+CARMENES+HIRES+APF+PFS+UVES spectra. \textbf{Bottom:} Periodogram of the residuals after the subtraction of the 143 days period signal with a double sinusoidal model including jitter terms and a linear trend.}
    \label{fig:Detect_allSignals_Halpha_Jitter}
\end{figure}

To complement this analysis, we introduce one jitter term for every single spectrograph in the double sinusoidal model and we change the independent term ($A_{3}$) for a linear trend term ($A_{3}+A_{4} \cdot t$), which leads to a very similar pre-whitening process shown in the Fig. \ref{fig:Detect_allSignals_Halpha_Jitter}. In this case, the error re-calculation associated with the jitter terms produce that the second signal to be detected shifts to 177 days with a FAP close to the 1\% level, and may also be related with differential rotation. The forest of peaks around the rotation period in the residuals is similar to the one shown in the bottom panel of Fig. \ref{fig:Detect_allSignals_Halpha}, giving us a rotation range between 130 and 180 days.

\subsection{Ca II HK index}

\begin{figure} 
	\includegraphics[width=\columnwidth]{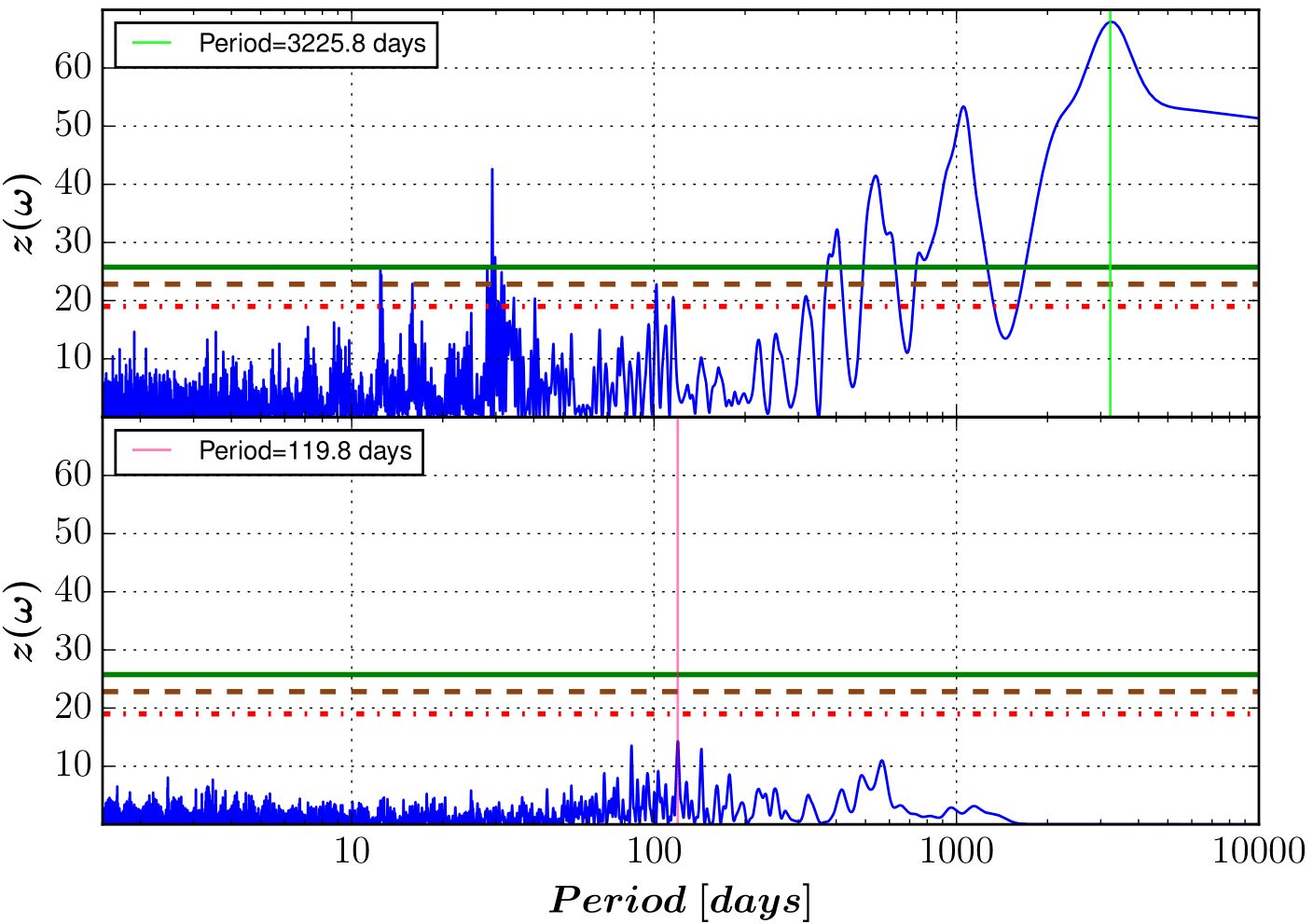} 
    \caption{\textbf{Top:} Periodograms of the time-series of CaHK for HIRES-Blue+HARPS-Pre2015+APF+HARPS-Post2015+HARPSN spectra. \textbf{Bottom:} Periodogram of the residuals after the subtraction of the 3225.8 days period signal.}
    \label{fig:Detect_allSignals_CaHK}
\end{figure}

\begin{figure} 
	\includegraphics[width=\columnwidth]{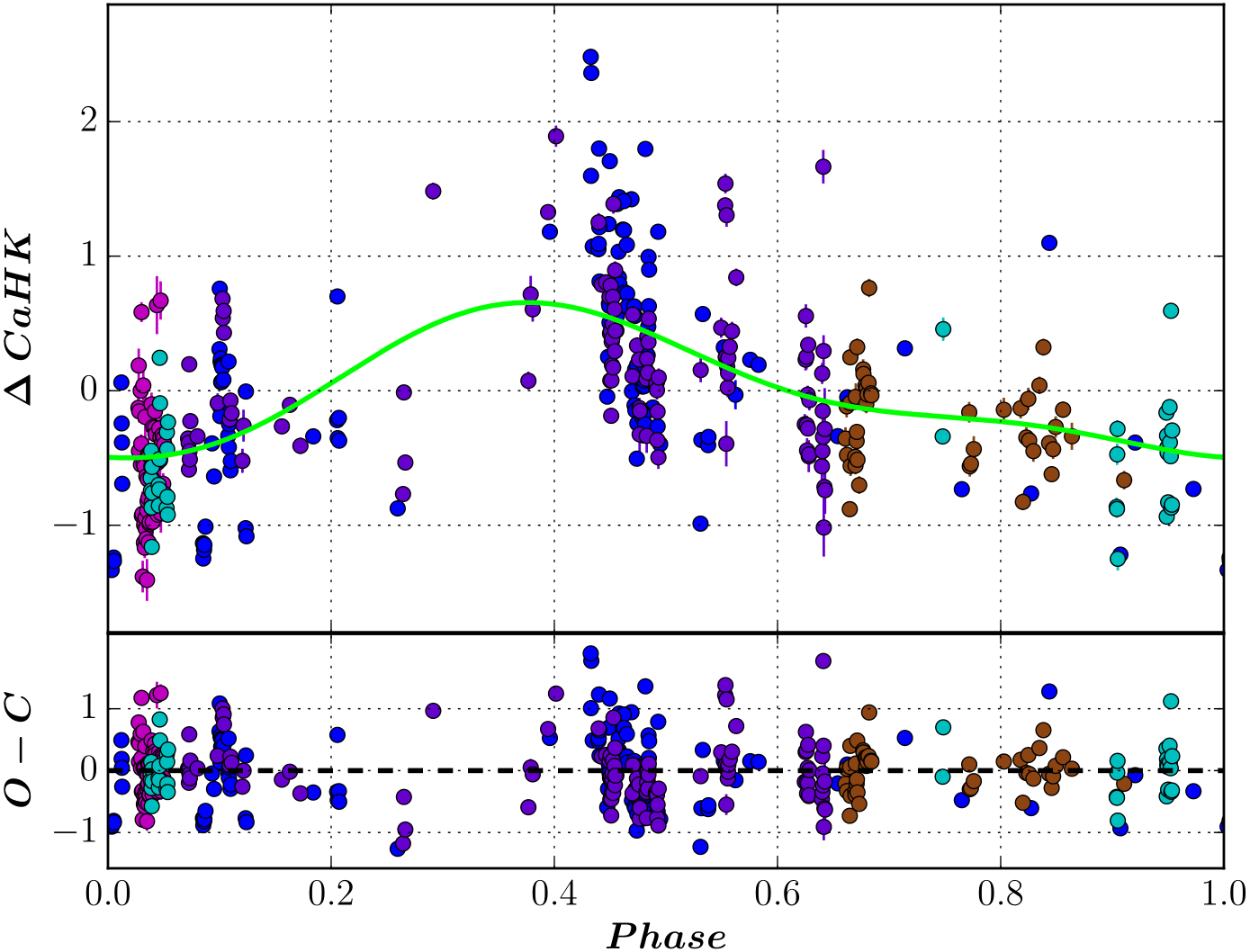} 
    \caption{Phase-folded curve of the CaHK time-series using the 3226 days period. Each spectrograph has been represented with a different color, following the legend in Fig. \ref{fig:Compact_Halpha_Smw_Na}. The green line represents the best double-sinusoidal fit found by the MPFIT routine.} 
    \label{fig:Periodogram_and_Fit_Smw}
\end{figure}

In the time-series of the CaHK index, the blue arm spectra of UVES, where the calcium lines are located, was not available. The CARMENES wavelength coverage did neither include the Ca II H\&K spectral range, and therefore it was not used. We also omitted PFS due to the high noise in that wavelength range. Therefore, we had 385 measurements of this index coming from HARPS, HARPS-N, HIRES and APF, with an average of 4.63, mean error of 0.06, and RMS of 0.6. Owing to the new continuum filters introduced in this work, we did not use the Mount Wilson calibration \citep{Vaughan1978} for this index. Using the four time-series with their respective offsets and without the Mount Wilson calibration, we first detected a 3225.8-day signal that remains stable after the trend subtraction, as it is shown in Fig. \ref{fig:Detect_allSignals_CaHK}. This long-period signal may be related to a long-term activity cycle in the star. It has a FAP level above the 0.1\%  and its fitted by the double sinusoidal shown in Fig. \ref{fig:Periodogram_and_Fit_Smw} that includes jitter terms and has an amplitude of 0.5 $\pm$ 0.4. When we subtract this model, we obtain a 120-day period signal with very low significance that seems to be dependent on the model used to subtract the long-term signal. Depending on the use of jitter terms and trend correction, we obtain different peaks in the range of $\sim$ 80 to 200 days with a similar FAP, so we could not ensure that any of the signals are indeed stellar activity signals. We also could not find a clear signal associated with the expected rotation in the analysis of the CaHK index.

\subsection{Na I D index}

\begin{figure} 
	\includegraphics[width=\columnwidth]{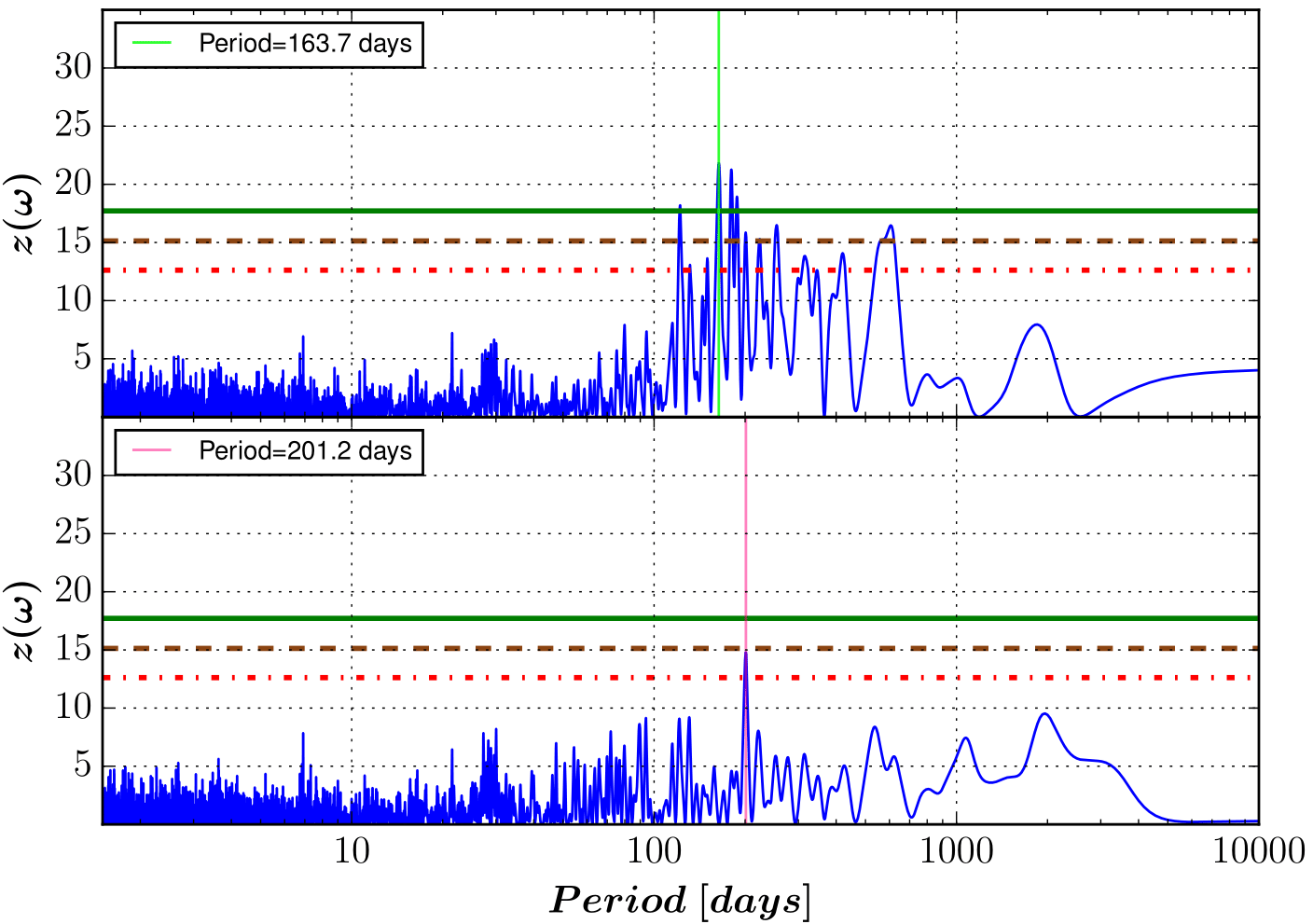} 
    \caption{Periodograms with the detected signals in the time-series of NaD for HARPS-Pre2015+PFS+APF+CARMENES+HARPS-Post2015+HARPSN spectra using the pre-whitening technique.}
    \label{fig:Detect_allSignals_NaID}
\end{figure}

\begin{figure} 
	\includegraphics[width=\columnwidth]{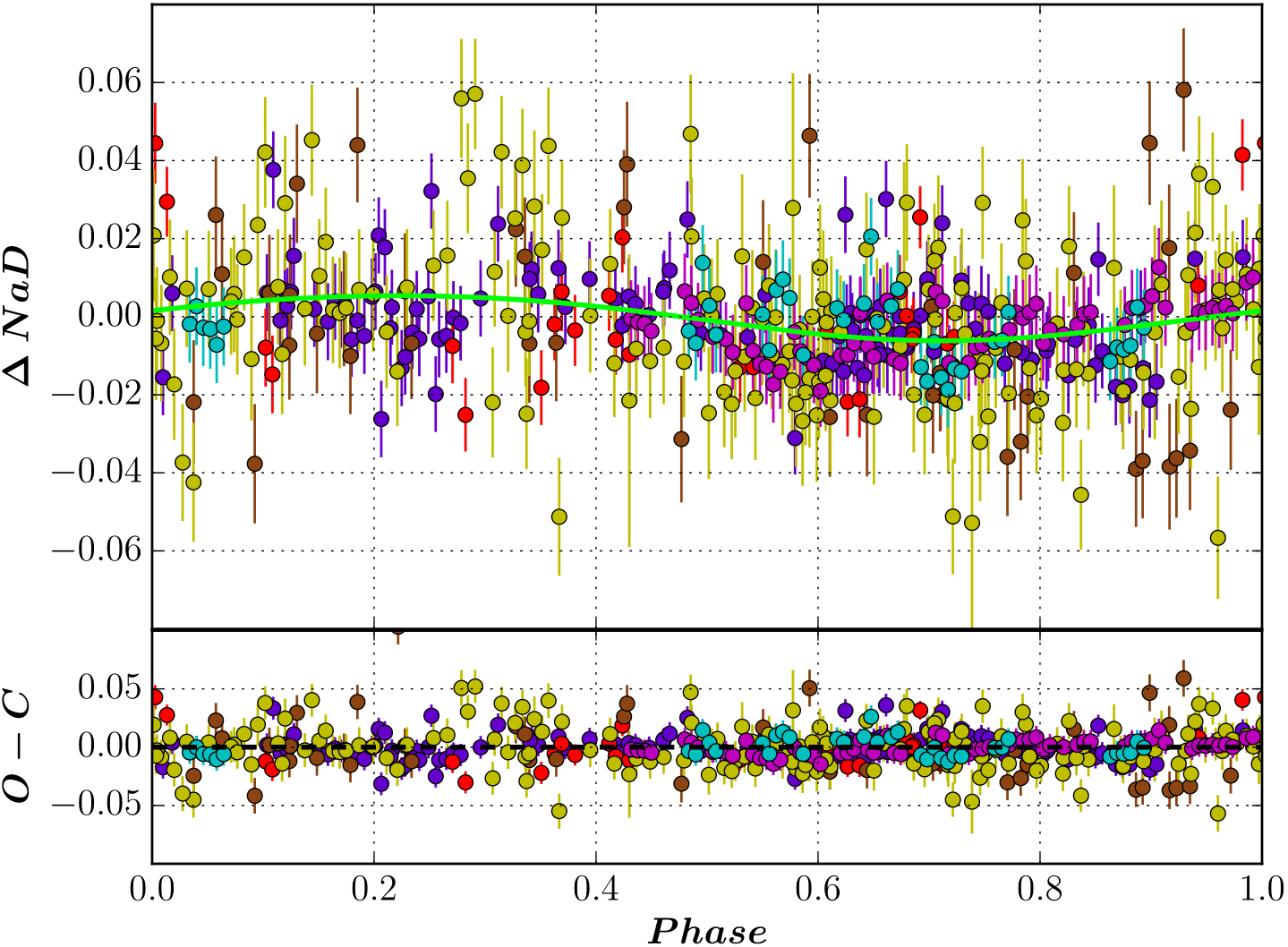} 
    \caption{Phase-folded curve of the NaD time-series using the 164 days period. Each spectrograph has been represented with a different color, following the legend in Fig. \ref{fig:Compact_Halpha_Smw_Na}. The green line represents the best double-sinusoidal fit found by the MPFIT routine.}
    \label{fig:Periodogram_and_Fit_Na}
\end{figure}

The time-series measurements of the Na I D index do not include HIRES data because we could not get a reliable wavelength calibration for the echelle orders that contains the core lines and the continuum regions. We also avoid using UVES due to the lower SN ratio in those orders. This leaves 453 measurements with an average of 0.19, mean error of 0.01, and RMS of 0.02.  When we treat this time-series and combine the two blocks of instruments we obtain the first periodogram shown in Fig. \ref{fig:Detect_allSignals_NaID}. We found a signal at 164 days surrounded by a forest of peaks similar to the one found in H$\alpha$ that could be associated with the rotation of the star. The difference in period with respect to the detected signal in the H$\alpha$ index suggests that this signal could be caused by differential rotation. In the Sun, for instance, the rotation period can vary from the equator (25 days) to the pole (35 days) in 40\%. From a sample of more than 24 000 active Kepler stars, \citet{Reinhold2013} found evidences of differential rotation within the 30\% of the equatorial rotation period in 77\% of the sample. In a more recient study, \citet{Aigrain2015} tested a blind hare-and-hounds exercise using 1000 simulated photometric light curves, and found little correlation between the reported and simulated values of the differential rotation, indicating that this detection in single light curves must be treated with caution. With a spectroscopic study like ours, using time-series from different activity indicators, we gain reliability with a detection of the same structure around the rotation period in two of the time-series. In this case, the variation from the original period measured in H$\alpha$ to the one measured in Na I D is only 15\%. This signal has an amplitude of 0.0070 $\pm$ 0.0008 and its FAP grows near the 0.1\%. When we subtract this signal with the double sinusoidal model shown in Fig. \ref{fig:Periodogram_and_Fit_Na} that includes jitter terms, the rest of the peaks remain with higher FAP values than 1\%, and they may be caused by the offsets between spectrographs, so no more clear information was extracted from this index. 

\subsection{Full width half maximum}

\begin{figure} 
	\includegraphics[width=\columnwidth]{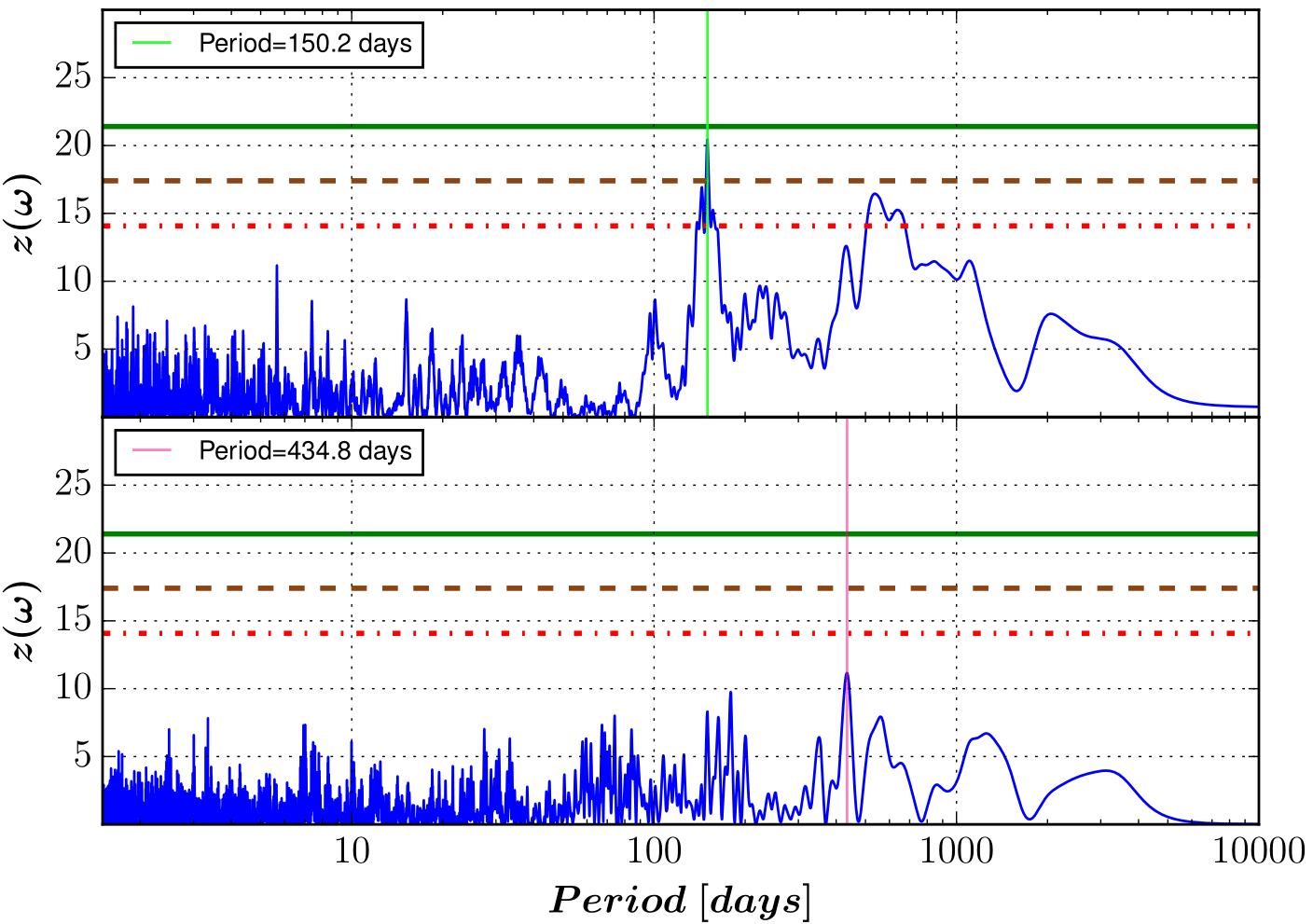} 
    \caption{\textbf{Top:} Periodograms of the time-series of FWHM for HARPS-Pre2015+CARMENES+HARPS-Post2015+HARPSN spectra. \textbf{Bottom:} Periodogram of the residuals after the subtraction of the 150 days period signal.}
    \label{fig:Detect_allSignals_FWHM}
\end{figure}

\begin{figure} 
	\includegraphics[width=\columnwidth]{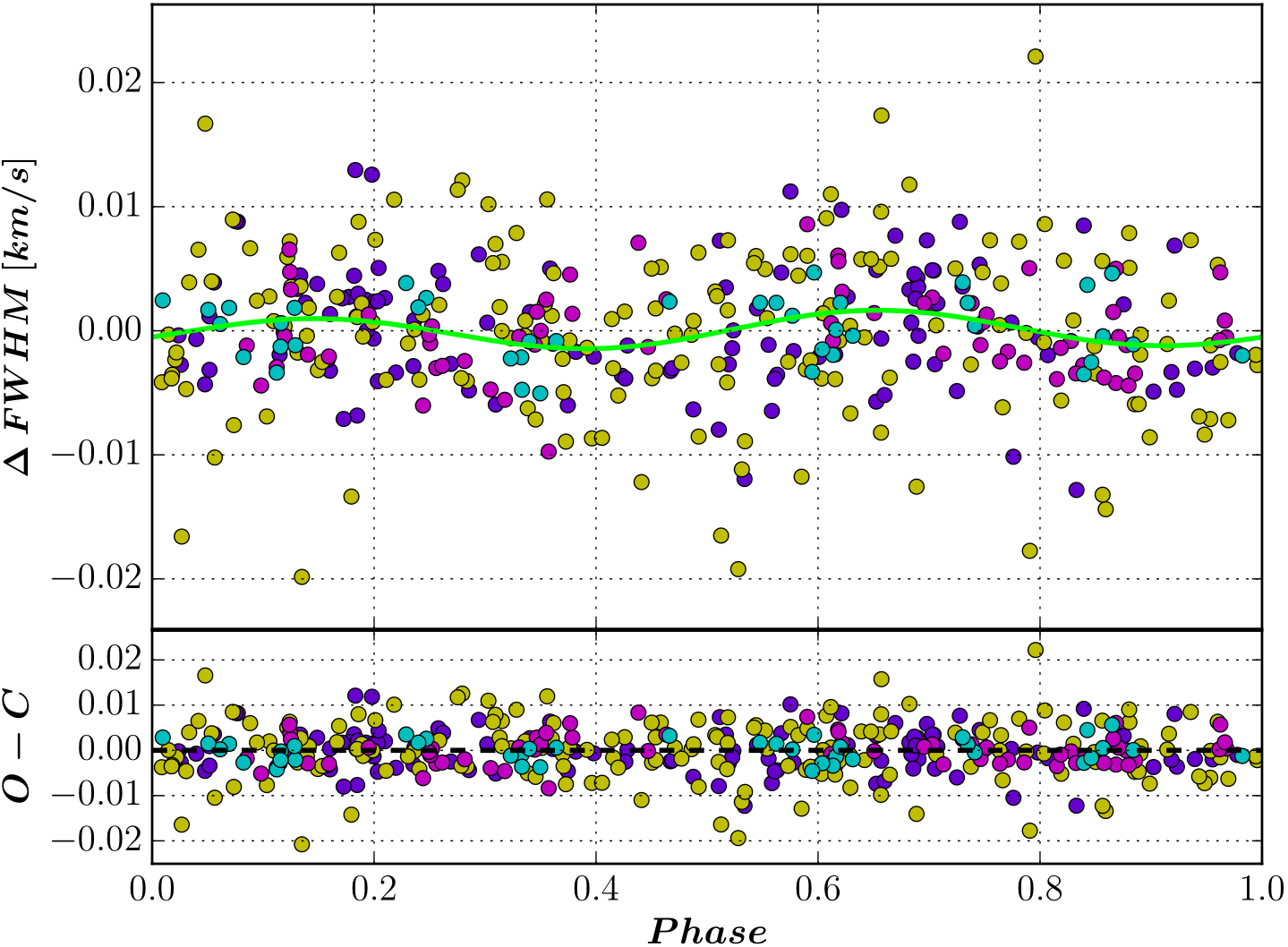} 
    \caption{Phase-folded curve of the FWHM time-series using the 150 days period. Each spectrograph has been represented with a different color, following the legend in Fig. \ref{fig:Compact_Halpha_Smw_Na}. The green line represents the best double-sinusoidal fit found by the MPFIT routine.}
    \label{fig:Periodogram_and_Fit_FWHM}
\end{figure}

Finally, the time-series of the FWHM, consists of 393 measurements, with an average of 4.52 km/s, mean error of 0.00005 km/s, and RMS of 0.006 km/s. We first apply a trend correction to the HARPS-Pre2015 values due to a focus drift problem. We also noticed a highest dispersion in the CARMENES values (see Fig. \ref{fig:Compact_Halpha_Smw_Na}) that may be related with the lack of weights per order in this spectrograph (the CCFs that we have used were already built as a one average function). The combination of the three spectrographs for which we have a CCF leads to a tentative detection of the rotation period at 150 days with a FAP level close to the 0.1\%, as it is shown in Fig. \ref{fig:Detect_allSignals_FWHM}. In this case, the signal its fitted by the double sinusoidal shown in Fig. \ref{fig:Periodogram_and_Fit_FWHM} with an amplitude of 0.00343 $\pm$ 0.00006 km/s. After subtracting this first peak with a double sinusoidal including jitter terms and a global trend, the remaining peaks do not exceed the 10\% level of FAP, making it difficult to establish a clear origin for them.

\subsection{Photometry}

We complement our spectroscopic analysis using the time-series of V-band photometric measurements. A higher number of data points are available (1390) compared to the spectroscopic dataset (619 as maximum), with an average of 9.5 mag, mean error of 9.2 mmag, and RMS of 15.4 mmag. We begin by analysing the largest dataset (ASAS), combining the  ASAS-S and ASAS-N time-series.

\begin{figure} 
	\includegraphics[width=\columnwidth]{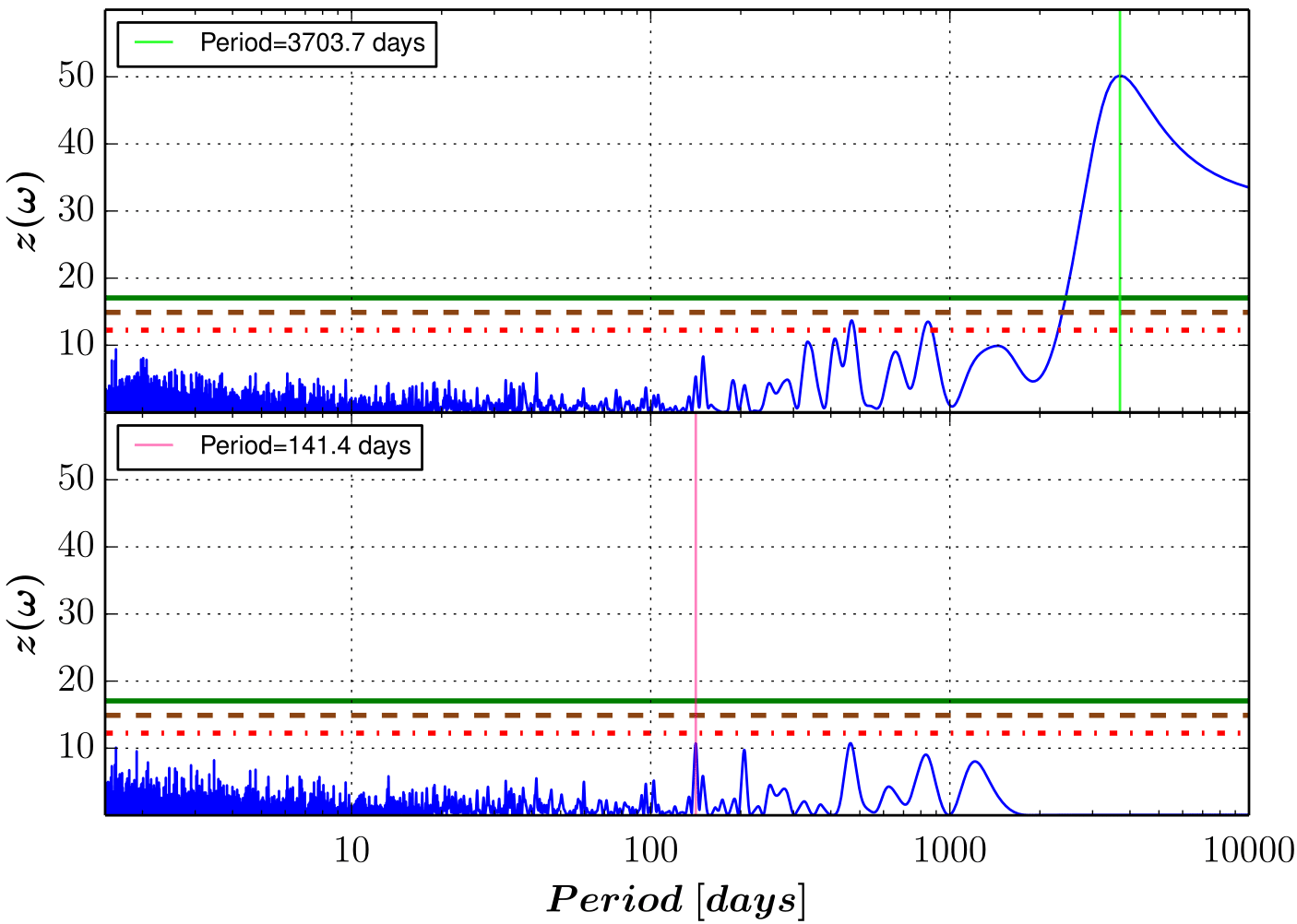} 
    \caption{\textbf{Top:} Periodogram of the time-series of ASAS-S+ASAS-N $m_{\rm V}$. \textbf{Bottom:} Periodogram of the residuals after the subtraction of the 3703.7 days period signal.}
    \label{fig:Detect_allSignals_ASAS}
\end{figure}

As it is shown the top panel in Fig. \ref{fig:Detect_allSignals_ASAS}, we find the first signal at 3703.7 days, which may be related to a long-term activity cycle in the star. After the subtraction of this signal with a very high amplitude (0.012 $\pm$ 0.004 mag), the rest of the peaks in the periodogram remain under the 10\% level of FAP, with the rotation period at 141 days being the second signal in amplitude. The addition of the ASAS-SN dataset produces a shift in the peak of the long-period signal to 3846 days, increasing its amplitude, and also producing an increase in the FAP levels from the bootstrapping.

\begin{figure} 
	\includegraphics[width=\columnwidth]{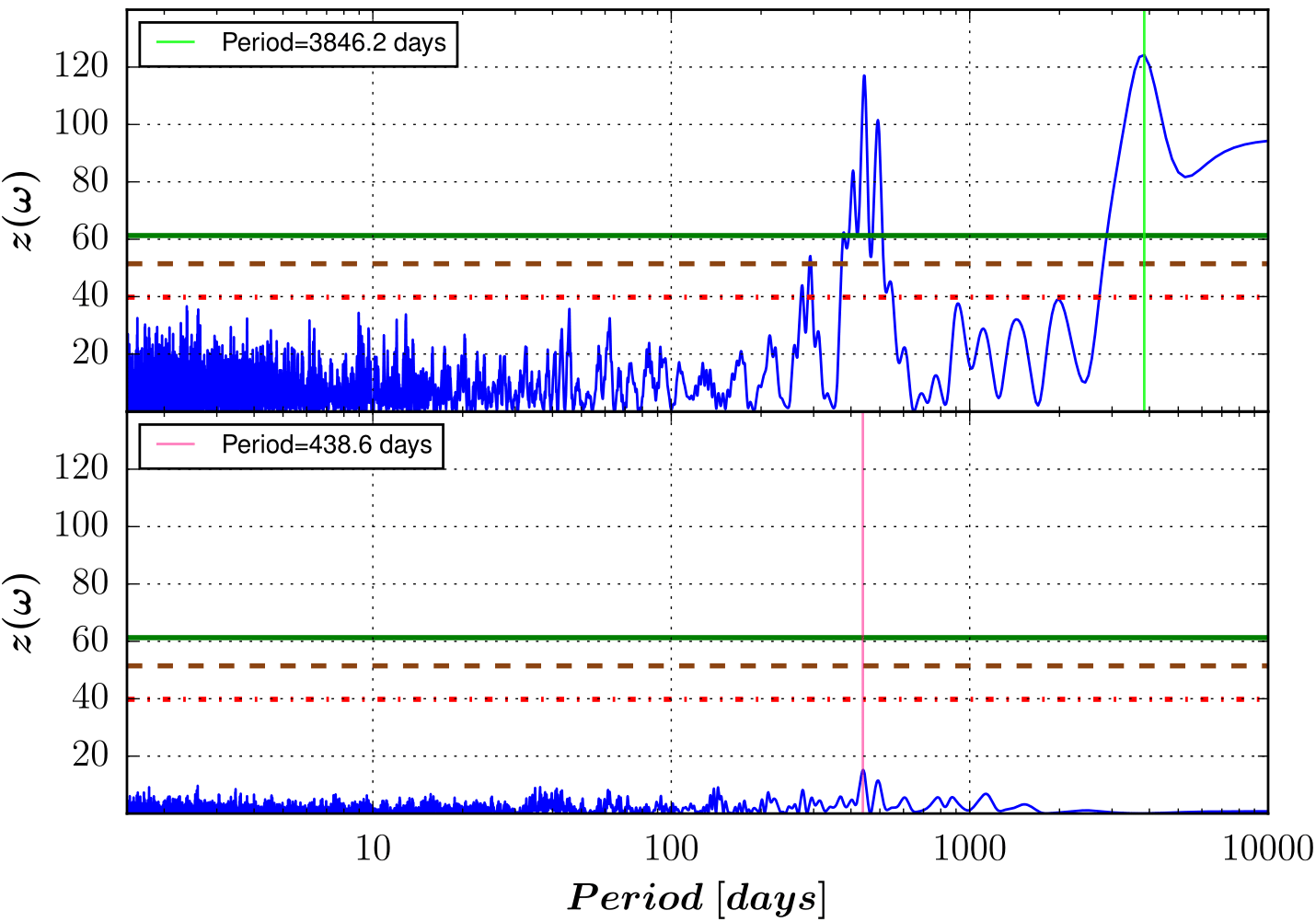} 
    \caption{\textbf{Top:} Periodogram of the time-series of ASAS+AAVSO+SNO $m_{\rm V}$. \textbf{Bottom:} Periodogram of the residuals after the subtraction of the 3846 days period signal.}
    \label{fig:Detect_allSignals_Phot}
\end{figure}

When we add the AAVSO and SNO datasets and determine the offsets using ASAS as reference, we recover the long-term activity cycle signal at 3846 days signal present in the complete ASAS dataset as it shows the first periodogram of Fig. \ref{fig:Detect_allSignals_Phot}, where the bootstrapping process to obtain the FAP levels was done omitting the ASAS-SN data. The double sinusoidal model that includes jitter terms and a linear trend presents an amplitude of 0.012 $\pm$ 0.006 mag, and its subtraction produces the periodogram shown in the bottom panel of Fig. \ref{fig:Detect_allSignals_Phot}, where the peak with the highest amplitude is a signal at 438.6 days with very low significance.

\begin{figure} 
	\includegraphics[width=\columnwidth]{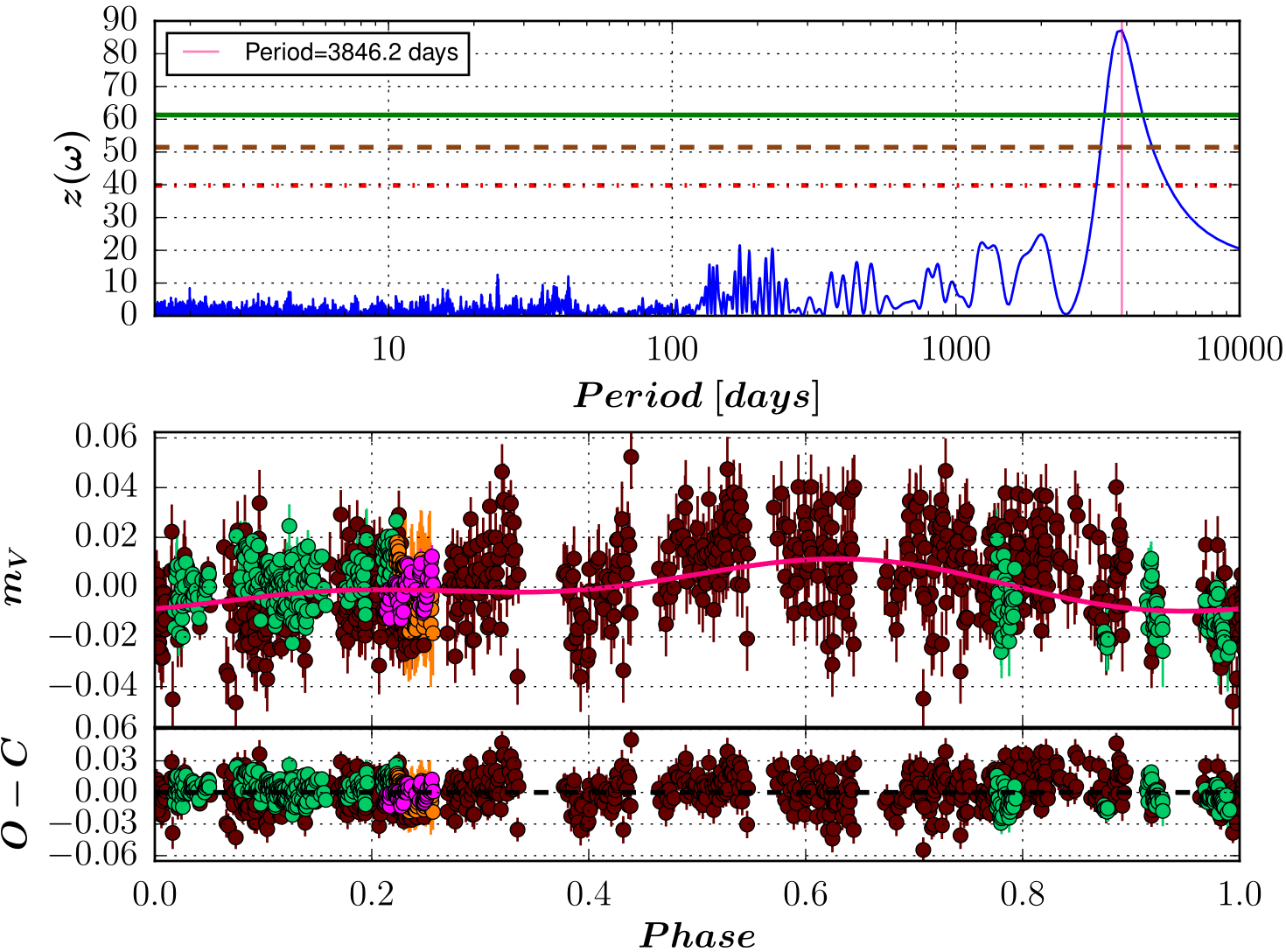}
    \caption{\textbf{Top:} Periodogram of the time-series of ASAS+AAVSO+SNO+FCAPT-RCT $m_{\rm V}$ after the subtraction of the 204.5 days period signal. \textbf{Bottom:} Phase-folded curve of the $m_{\rm V}$ time-series using the 3846.2 days period. Each instrument has been represented with a different color, following the legend in Fig. \ref{fig:Compact_Halpha_Smw_Na}. The pink line represents the best double-sinusoidal fit found by the MPFIT routine.}
    \label{fig:Periodogram_and_Fit_Phot_All}
\end{figure}

The addition of the FCAPT-RCT dataset (see Fig. \ref{fig:Periodogram_and_Fit_Phot_All}) creates a broad and signal at 204.5 days as the most significant one. After the subtraction of this first signal, we recover the long-term activity cycle signal at 3846.2 days obtained in previous time-series with an amplitude of 0.009 $\pm$ 0.008 mag. This signal is shown in Fig. \ref{fig:Periodogram_and_Fit_Phot_All}, where the periodicity of the cycle is even more clear than in the time-series of the CaHK index due to the higher number of points. The FAP values come from the bootstrapping process carried out omitting the ASAS-SN and FCAPT-RCT values.

The problem with the FCAPT-RCT dataset is the underestimation in the $m_{\rm V}$ errors, which produces the high amplitude signal at 204.5 days in the periodogram along with higher FAP values from the bootstrapping process that can make the long-term activity cycle signal be confused with noise because it does not reach the 10\% level of FAP. When we apply a jitter term to this dataset and recalcute the errors, we recover the long-term signal with its expected amplitude. It is also important to take into account the time gap of $\sim$ 8 years between the FCAPT and RCT datasets, which can affect the results due to a bigger uncertainity in the offset between those two datasets. The same happens between the ASAS-S+ASAS-N and ASAS-SN dataset, with a gap of more than 1 year between them, but in this case the difference is not so remarkable. We maintain the ASAS-SN time-series in the analysis because is needed to obtain the offset values using the time windows methodology and the FCAPT-RCT time-series because it increases the $m_{\rm V}$ amplitude of the long-term activity cycle signal.

In the separate analysis of Montsec and MEarth time-series from different photometric filters, we do not detect any significant signal that could be attributed to rotation or a long-term activity cycle, although MEarth has proven to be capable of detecting rotation periods \citet{Newton2016a}.

\subsection{Chromatic index and Bisector span}

We also did an additional analysis using the time-series of the chromatic index (CRX) that contains 216 measurements taken by CARMENES in a 2-year time span. This activity indicator was defined by \citet{Zechmeister2018} and it serves to measure the RV-wavelength dependence. The CARMENES pipeline correlates these two quantities along all the echelle orders and then fits a first-order-polynomial whose slope is taken as a measurement of the CRX.

In the time-series of this index, we first found a 10000 day-signal that dissapears after a trend correction. This hints to the presence of a change in the level of activity on time scales much larger than the range of our observations. After the correction, the FAP of the most significant signal is greather than 10\%, so we could not find anything relevant in this time-series.

We also analysed the bisector span time-series (BIS), an index that comes from the slope of the polynomial that fits the centroid of the CCF at different heights \citep{Queloz2001}. The BIS time-series is composed by 116 measurements of HARPS-N, 31 of HARPS-N and 186 of CARMENES. This index measures the distortion of the CCF under the presence of stellar spots and plages. This distortion is lower in fast rotators and low activity stars. As for the CRX index, we could not find any significant signal in the analysis of the BIS time-series.

\section{Discussion}

\label{sec:Discussion}

Combining the rotation period from H$\alpha$ and FWHM time-series with weights according to their FAP level we obtain a final average value of 145 $\pm$ 15 days. This 10\% error comes from the FWHM of a gaussian model that fits the forest of peaks around the 145-day signal and takes into account the uncertainty in the latitude of the active regions that are producing this signal. This means that Barnard's Star is among the main sequence stars with lowest rotation known to date, above the M-stars average periods \citep{Mascareño2018b,Newton2016a}. This also suggests that Barnard's age matches the age of the local thick disk \citep{Newton2018}.

Differential rotation may be responsible for the different signals found in between 130 to 180 days, as a consequence of the presence of active regions at different latitudes of the stellar surface. This phenomenon has not been fully understood for stars from all spectral types, but especially for M-dwarfs. \citet{Reinhold2015} confirmed a relation between rotation period and differential rotation predicted by \citet{Reiners2003} including M-type stars in their sample. Although this relation has only been proven for stars with P$_{\rm min} <$ 50 days, we obtain a value for Barnard's Star of $\alpha$=(Pmax-Pmin)/Pmax=0.278 that matches the M-stars values present in this study. Taking into account that differential rotation is more evident in slow rotators, we conclude that our estimation is consistent with the theoretical prediction for differential rotation.

\begin{figure} 
	\includegraphics[width=\columnwidth]{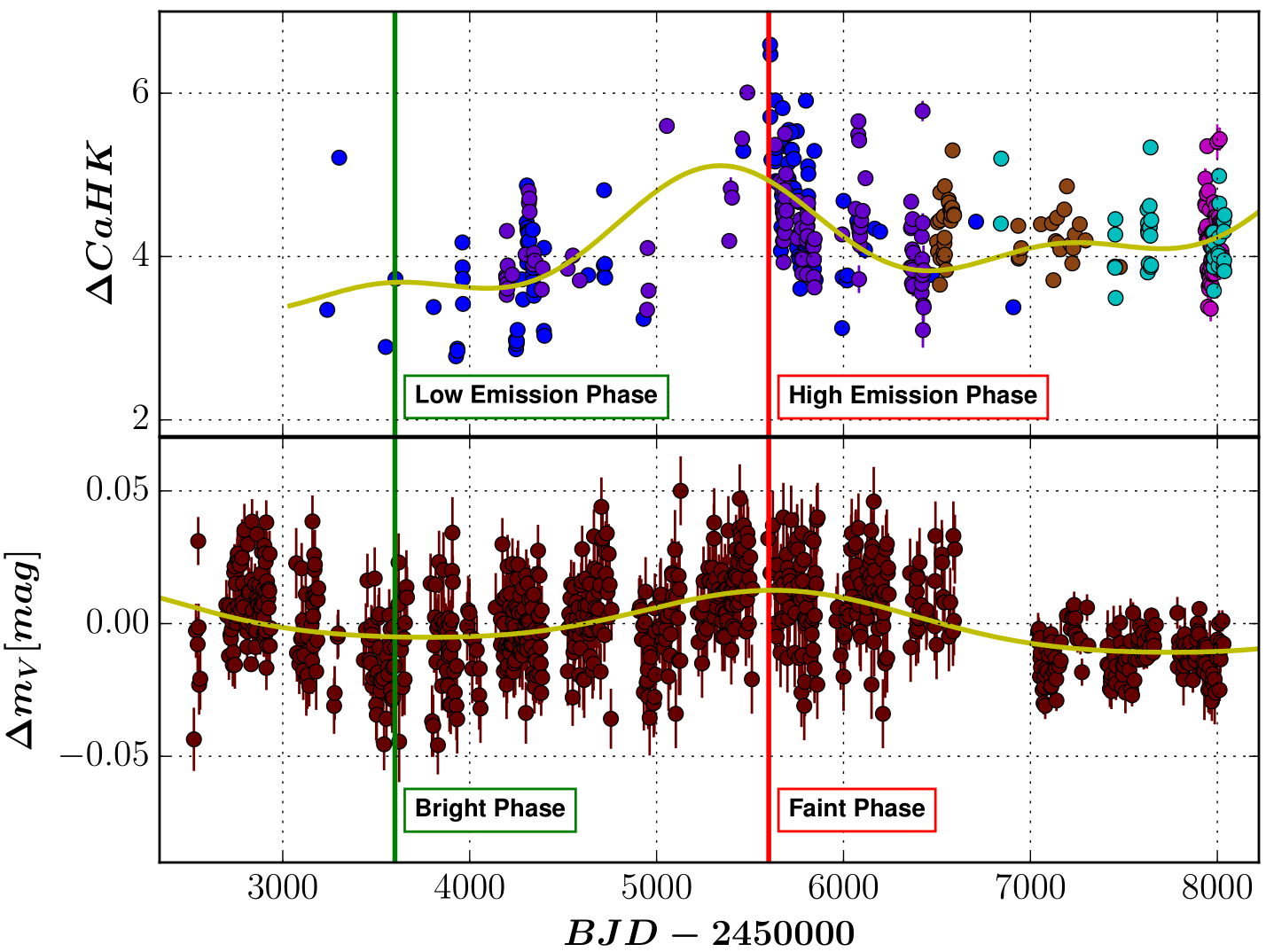} 
    \caption{\textbf{Top:} Time-series of CaHK for HIRES-Blue+HARPS-Pre2015+APF+HARPS-Post2015+HARPSN spectra. The beginning of the low emission phase is marked in green and the beginning of the high emission is marked in red. \textbf{Bottom:} Time-series of ASAS-S+ASAS-N $m_{\rm V}$. The beginning of the bright phase is marked in green and the beginning of the faint is marked in red. In both panels, the yellow line represents the best double-sinusoidal fit found by the MPFIT routine.}
    \label{fig:Cycle}
\end{figure}

We detected two similar long period signals, in the Ca II H\&K and m$_{\rm V}$ time-series. The two signals show similar periodicities, both compatible with the length of a solar-like cycle. When we compare the two series side by side (see Fig.~\ref{fig:Cycle}) we can see even more similarities. Not only their periods are compatible, but the dates of their maxima and minima are virtually the same, hinting a common underlying phenomenon. The combination of the two series gives us coverage along two full phases of the signal, pointing at a cyclic nature. We can interpret this variability as the footprint of a magnetic cycle of 10 $\pm$ 2 years, which is not expected for a completely convective star like Barnard's Star \citep{Chabrier2006,Wargelin2017}.

It is interesting to note that the position of the maximum emission phase coincides with the position of the faintest phase of the star and the minimum emission phase coincides with the brightest phase of the light-curve. Given the low level of chromospheric and X-ray emission of Barnard's Star \citep{Passegger2018}, this behaviour is opposite to the solar case, and to most old FGK stars \citep{Radick1998}. It would be compatible with a spot-dominated stellar surface, typical of active FGK stars. In active stars, spots dominate the brightness changes, while plages would dominate chromospheric and X-ray emission. The situation is similar to what \citet{Wargelin2017} found for the case of Proxima, when comparing \textit{V}-band photometry to X-ray and UV emission. Despite being old, Proxima remains quite active \citep{Pavlenko2017}, which made it natural to put in on the ``active stars'' category. The case of Barnard is quite different, as the star shows very low levels of chromospheric and X-ray emission. This could hint at late M-dwarfs keeping the ``active star'' behaviour, and remaining spot dominated, even after their chromospheric and X-ray emission reach extremely low levels.

Given our short baseline, the exact period and long-term behaviour are still complicated to asses. Further monitoring spectroscopic and photometric would be needed to better characterize it.

The amplitude of the rotation period and long-term activity cycle signals are shown in Table \ref{tab:Semi-amplitudes}.

\begin{table}
\centering
	\caption{Semi-amplitude of the isolated signals from the four spectroscopic indexes and the photometric magnitude.}
	\label{tab:Semi-amplitudes}
	\begin{tabular}{lcc}
		\hline
		Index & \textit{P} [days] & Semi-amplitude  \\
		\hline
		\multirow{2}{*}{H$\alpha$}    & 143.68  & 0.00523 $\pm$ 0.00001      \\
		                              & 149.03  & 0.00305 $\pm$ 0.0001       \\
		\multirow{1}{*}{CaHK}         & 3225.81 & 0.5     $\pm$ 0.4          \\
		\multirow{1}{*}{NaD}          & 163.67  & 0.0070  $\pm$ 0.0008       \\
		\multirow{1}{*}{FWHM}         & 150.15  & 0.00343 $\pm$ 0.00006 km/s \\
		\multirow{1}{*}{$m_{\rm V}$}  & 3846.15 & 0.009   $\pm$ 0.008 mag    \\
		\hline
	\end{tabular}
\end{table}

Applying the Mount Wilson calibration to the S-index of all spectrographs by the following expression: 

\begin{equation}
  S_{mw}=\alpha \cdot S +\beta
  \label{eq:Smw_index}
\end{equation}

\noindent where $\alpha=1.111$, $\beta=0.0153$ \citep{Lovis2011} and $S$ is calculated with the original passbands, we can use its mean value $<S_{mw}>$ to calculate the level of chromospheric activity log$_{10} ($R$_{\rm HK}^{'})$ as \citep{Noyes1984}:

\begin{equation}
  log_{10}(R'_{HK})=log_{10}\left((1.34 \cdot 10^{-4}) \cdot C_{cf} \cdot <S_{mw}> -R_{phot}\right)
  \label{eq:logRhk}
\end{equation}

\noindent where C$_{\rm cf}$ is a conversion factor to correct the flux variations in the continuum passbands and also to normalize to the bolometric luminosity, that is defined as \citep{Mascareño2015}:

\begin{equation}
  log_{10}(C_{cf})=-0.443-0.645 \cdot (B-V) -1.270 \cdot (B-V)^{2}+0.668 \cdot (B-V)^{3}
  \label{eq:Ccf}
\end{equation}

\noindent and R$_{\rm phot}$ is the photospheric contribution to the calcium core lines \citep{Hartmann1984} that we need to get rid of in order to measure only the chromospheric contribution:

\begin{equation}
  log_{10}(R_{phot})=(1.48 \cdot 10^{-4}) \cdot e^{-4.3658 \cdot (B-V)}
  \label{eq:Rphot}
\end{equation}

Eq. ~(\ref{eq:logRhk}) gives a chromospheric activity level of log$_{10}(R^{'}_{\rm HK})$=-5.82 $\pm$ 0.08 using our $S_{mw}$ measurements of Gl 699 that is in good agreement with the values of -5.69 \citep{Astudillo2017} and -5.86 \citep{Mascareño2015} from the literature. If we use the relation between the chromospheric activity level and induced RV semi-amplitude found by \citep{Mascareño2017,Mascareño2018b}, we get an induced semi-amplitude of $K=0.67 ^{+0.28}_{-0.20}$ m/s, which give us an upper limit of 0.95 m/s that marginally falls on the detection limit for most of the current instrumentation dedicated to RV searches \citep{Pepe2014}. 

\begin{figure} 
	\includegraphics[width=\columnwidth]{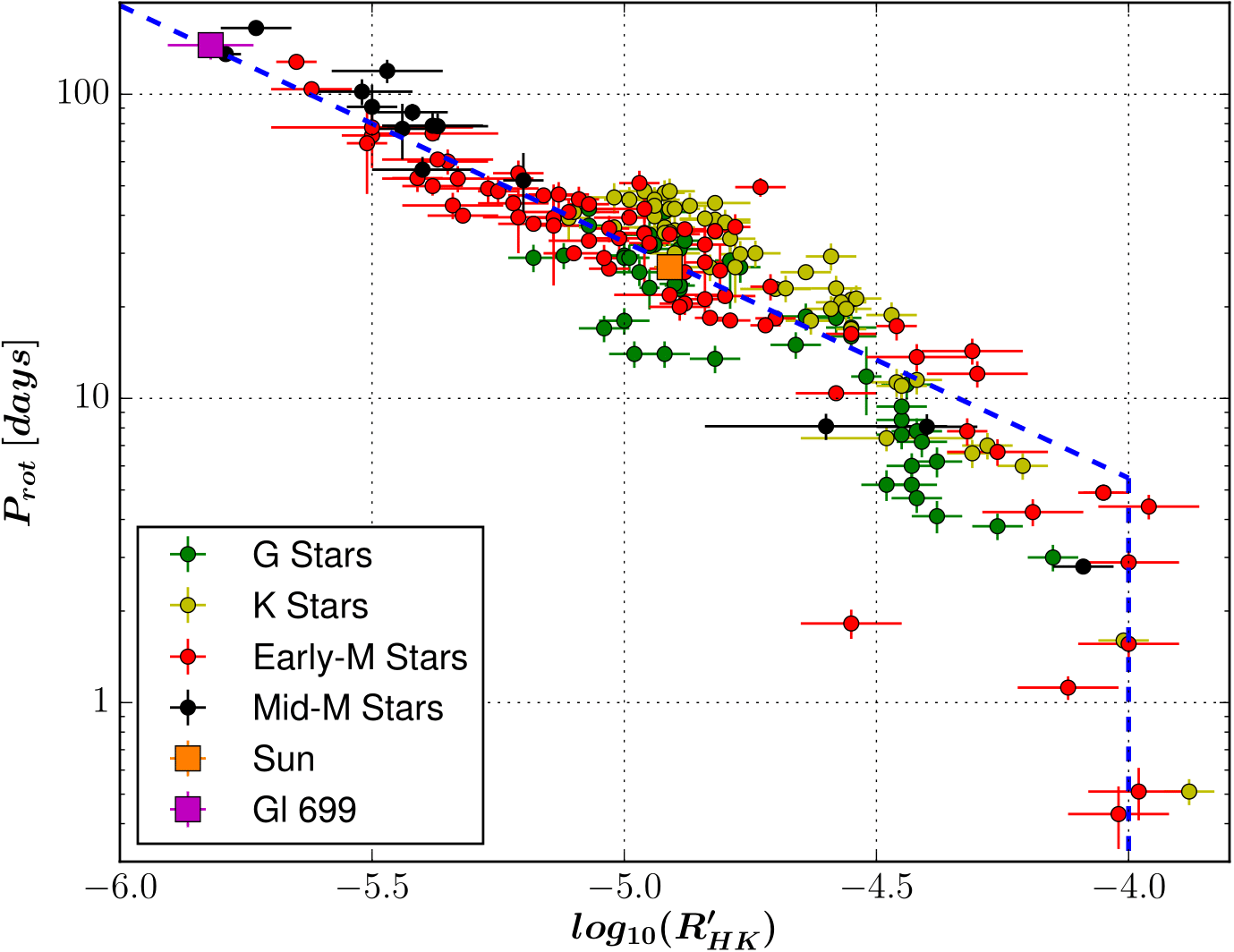} 
    \caption{Rotation period and chromospheric activity level log$_{10} R_{HK}^{'}$ of a G, K, Early-M (M0 to M3) and Mid-M (M4 to M6) stars sample from the literature, including the values obtained in this paper for Gl 699. The blue line represents the relation obtained by \citet{Mascareño2018b} for M stars.}
    \label{fig:Prot_logRhk}
\end{figure}

In the last years, several groups have studied the rotation periods of a large sample of stars \citep[see e.g.][]
{McQuillan2014,Newton2016a,Díez-Alonso2019}. We have selected a sample from \citet{Mascareño2015,Mascareño2016,Mascareño2018a,Mascareño2018b} and \citet{Astudillo2017}, to see how the values of the rotation period and the level of chromospheric activity obtained for Barnard's Star fit into the relation found by \citet{Mascareño2016}:

\begin{equation}
  log_{10}(P_{rot})=A + B \cdot log_{10}(R^{'}_{HK})
  \label{eq:Prot-logRhk}
\end{equation}

\noindent where $A=-2.37\pm0.28$ and $B=-0.777\pm0.054$ for M-type stars with a log$_{10}(R^{'}_{\rm HK}) \leq$ -4.1 \citep{Mascareño2018b}. As it is shown in Fig. \ref{fig:Prot_logRhk}, our rotation period fits very well into this theorical prediction (the rotation period value given by this relation is 142 days).

Also previous studies \citep{Pizzolato2003,Wright2011,Reiners2014,Wright2016} have investigated the relation between the X-ray emission and the rotation rate of stars. Using the relation found by \citet{Wright2011} we estimate a rotation period of 132 days, which is consistent with our result.

In the photometric time-series, we found the rotation period at 141 days period in the ASAS subset of data (that spans about 15 years), although with low significance (FAP $<$ 10\%, similar to the level obtained by \citet{Benedict1998} with the HST data). A period of $\sim$200 days is also present when we add the FCAPT-RCT dataset. This dataset has a time coverage similar  to ASAS (14 years), but with much sparse (about a third of the ASAS data, 348 to 836 epochs) and uneven gapped data, so that the offset could be responsible for the discrepancies in the photometric results, or less likely, differential rotation, as 180 days would be $\sim$30\% of the estimated 140 days rotation period.

The year-alias of a rotation period of 145d of happens at a period of about 240d \citep{Dawson2010} that is close to the planetary signal of 233d of Barnard b \citep{Ribas2018}. We computed the periodogram of the RV data using the Systemic console, including a linear trend term along with offset and jitter terms for each instrument as free parameters, which removes any possible long-term signal (of several years) that could be associated with either long-term activity or long-period planetary signals. The first periodogram after minimizing the linear term, offset and jitter values provides the strongest signal at 233d and secondary much less significant signals at 1d and 77d, but no significant signal around 145d. A real signal should appear at its original frequency and also at its two alias frequencies with a given significance depending on the level of noise \citep[see e.g.][]{Anglada-Escude2013}. The signal at 145d does not appear in the periodogram of the RV data. However, we performed several tests with the published set of RV time-series, trying to force a fit for the rotation signal (with both a sinusoid and a keplerian model), allowing the period to move in the range between 130d and 160d. After fitting and subtracting the stellar rotation signal, the 233d signal remains highly significant in the periodogram of the residuals. We note that 77d is about half of the estimated rotation period from activity indicators but we do not find any signal in the RV time-series at about 145d. After fitting and subtracting the 77d signal, the 233d signal still holds with high power. The combined fit of the 77d + 233d signals does not affect the final parameters of the planet reported in \citet{Ribas2018}. We conclude that the planetary signal at 233 days is not related to any possible rotation signal present in the RV data, which we have not been able to detect since it is probably much weaker than the RV precision of the data. The RV analysis is extensively discussed in \citet{Ribas2018} in the context of the activity signal associated with stellar rotation and the planetary signal associated with Barnard b.

\section{Conclusions}

\label{sec:Conclusions}

We have analysed the time variability of several spectroscopic indexes (H$\alpha$, CaII H\&K, NaI D, and CCF's FWHM) in a sample of 964 spectra of Barnard's Star taken with seven different spectrographs (HARPS, HARPS-N, CARMENES, HIRES, APF, PFS and UVES) in a time-span of 14.5 years. We also have used the available photometric time series of the star that forms a sample of 1390 measurements of photometric magnitudes coming from four different instruments (AAVSO, FCAPT-RCT, ASAS, and T90@SNO) in a time-span of 15.1 years. 

We have detected the rotation period signal in the H$\alpha$ and FWHM time-series at 143 and 150 days respectively, along with a tentative detection of differential rotation between 130 and 180 days appearing also in the NaD time-series. We determine the rotation period to be 145 $\pm$ 15 days for Barnard's Star. We also calculate a chromospheric activity level of log$_{10}(R^{'}_{\rm HK})$=-5.82 that indicates a very low stellar activity. Using an activity-rotation relation, we obtain an expected rotation period that is in good agreement with our determination, and an upper limit to the activity induced RV signal associated to rotation of 1 m/s. Also, the low X-ray activity of the star supports our determination of the stellar rotation period.

In the CaHK and $m_{\rm V}$ time-series we find evidence of a long-term activity cycle in 3226 days and 3846 days respectively, which is consistent with previous estimates of magnetic cycles from photometric time-series in other M stars with similar activity levels. We then derive a long-term activity cycle of 3800 $\pm$ 600 days for Barnard's Star.

We found no evidence that the signals detected in the chromospheric activity indicators are causing the RV signal detected by \citet{Ribas2018}.

\section*{Acknowledgements}

This work has been financed by the Spanish Ministry of Science, Innovation and Universities (MICIU) through the grant AYA2017-86389-P. B.T.P. acknowledges Fundaci\'on La Caixa for the financial support received in the form of a Ph.D. contract. J.I.G.H. acknowledges financial
support from the Spanish MICIU under the 2013 Ram\'on y
Cajal program MICIU RYC-2013-14875. A.S.M acknowledges
financial support from the Swiss National Science Foundation
(SNSF). The IAA-CSIC and UCM teams acknowledge support by the Spanish Ministry of Economy and Competitiveness (MINECO) through grants AYA2016-79425-C3-1-P, AYA2016-79425-C3-2-P, AYA2016-79425-C3-3-P, ESP2014-54362P, and ESP2017-87143R. I.R., J.C.M., M.P., and E.H acknowledge support from the Spanish MINECO and the Fondo Europeo de Desarrollo Regional (FEDER) through grant ESP2016-80435-C2-1-R, as well as the support of the Generalitat de Catalunya/CERCA program. G.A-E research is funded via the STFC Consolidated Grants ST/P000592/1, and a Perren foundation grant. The results of this paper were based on observations made with the Italian Telescopio Nazionale Galileo (TNG), operated on the island of La Palma by  the INAF-Fundaci\'on Galileo Galilei at the Roque de Los Muchachos Observatory of the Instituto de Astrof\'isica de Canarias (IAC); observations made with the HARPS instrument on the ESO 3.6-m telescope at La Silla Observatory (Chile); observations made with the CARMENES instrument at the 3.5-m telescope of the Centro Astron\'omico Hispano-Alem\'an de Calar Alto (CAHA, Almer\'ia, Spain), funded by the German Max-Planck-Gesellschaft (MPG), the Spanish Consejo Superior de Investigaciones Cient\'ificas (CSIC), the European Union, and the CARMENES Consortium members. This paper made use of the IAC Supercomputing facility HTCondor (\url{http://research.cs.wisc.edu/htcondor/}), partly financed by the Ministry of Economy and Competitiveness with FEDER funds, code IACA13-3E-2493. We are grateful to all the observers of the projects whose data we are using for the following spectrographs: HARPS (072.C-0488, 183.C-0437, 191.C-0505, 099.C-0880), HARPS-N (CAT14A\_43, A27CAT\_83, CAT13B\_136, CAT16A\_109, CAT17A\_38, CAT17A\_58), CARMENES (CARMENES GTO survey), HIRES (U11H, U11H, N12H, N10H, A264Hr, A288Hr, C168Hr, C199Hr, C205Hr, C202Hr, C232Hr, C240Hr, C275Hr, C332Hr, H174Hr, H218Hr, H238Hr, H224Hr, H244Hr, H257Hr, K01H, N007Hr, N014Hr, N024, N054Hr, N05H, N06H, N085Hr, N086Hr, N095Hr, N108Hr, N10H, N112Hr, N118Hr, N125Hr, N129HR, N12H, N12H, N131Hr, N131Hr, N136Hr, N141Hr, N145Hr, N148Hr, N14H, N157Hr, N15H, N168Hr, N19H, N20H, N22H, N28H, N32H, N50H, N59H, U014Hr, U01H, U023Hr, U027Hr, U027Hr, U030Hr, U052Hr, U058Hr, U05H, U064Hr, U077Hr, U078Hr, U07H, U082Hr, U084Hr, U08H, U10H, U115Hr, U11H, U12H, U131Hr, U142Hr, U66H, Y013Hr, Y065Hr, Y283Hr, Y292Hr), UVES (65.L-0428, 66.C-0446, 267.C-5700, 68.C-0415, 69.C-0722, 70.C-0044, 71.C-0498, 072.C0495, 173.C-0606, 078.C-0829), APF (LCES/APF planet survey) and PFS (Carnegie-California survey).

\begin{table*}
	\centering
	\caption{AAVSO contributions.}
	\label{tab:aavso}
	\begin{tabular}{lccccc} 
		\hline
		Observer code & Name &  Country &   Filters    & N$_{\rm exposures}$     &  N$_{\rm epochs}$   \\
		\hline
BJFB & John Briol	              & USA & \textit{V}                     & 334                & 13     \\
BLOC & Lorenzo Barbieri           & IT  & \textit{V}, \textit{H$\alpha$} & 523, 59            & (9)    \\
CIVA & Ivaldo Cervini             & CH  & \textit{V}, \textit{H$\alpha$} & 161, 70            & (12)   \\
DLM  & Marc Deldem                & FR  & \textit{V}                     & 2015               & 28     \\
DUBF & Franky Dubois              & BE  & \textit{BVRI}                  & 210, 233, 188, 124 & (31)   \\	
HBB  & Barbara Harris	          & USA & \textit{V}                     & 463                & 4      \\	
HMB  & Franz-Josef Hambsch        & BE  & \textit{V}                     & 2753               & 111    \\	
KCLA & Clifford Kotnik            & USA & \textit{V}, \textit{H$\alpha$ }& 867, 83            & (8)    \\
LJBE & Jean-Marie Lopez           & FR  & \textit{V}                     & 446                & 6      \\		
MMAE & Michael McNeely            & USA & \textit{V}                     & 2                  & (2)    \\	
OYE  & Yenal Ogmen                & CY  & \textit{V}                     & 416                & 1      \\	
PLFA & Luis P\'erez               & ES  & \textit{V}                     & 65                 & 1      \\   
RZD  & Diego Rodr\'iguez          & ES  & \textit{V}                     & 1                  & (1)    \\    
SFGA & Fabi\'an S\'anchez Urquijo & EC  & \textit{V}                     & 2                  & (2)    \\    	       	
		\hline
	\end{tabular}
		\begin{minipage}{2\columnwidth} 
{\footnotesize \textbf{Columns:} Observer initials and name, country code (USA=United States of America; IT=Italy; CH=Switzerland; FR=France; BE=Belgium; CY=Cyprus; ES=Spain; EC=Ecuador), filters, number of exposures and number of epochs. The parenthesis in the last column indicates that the datasets were not included in the final analysis due to high scattering or insufficient number of observations.}
\end{minipage}	
\end{table*}

\section*{Affiliations}

$^{\boldsymbol{1}}$ Instituto de Astrof\'isica de Canarias, E-38205 La Laguna, Tenerife, Spain\\
$^{\boldsymbol{2}}$ Universidad de La Laguna, Departamento de Astrof\'isica, E-38206 La Laguna, Tenerife, Spain\\
$^{\boldsymbol{3}}$ Instituto de Astrof\'isica de Andaluc\'ia, Glorieta de la Astronom\'ia 1, 18008, Granada, Spain\\
$^{\boldsymbol{4}}$ Observatoire Astronomique de l'Universit\'e de Gen\`eve, 1290 Versoix, Switzerland\\
$^{\boldsymbol{5}}$ Consejo Superior de Investigaciones Cient\'ificas, E-28006 Madrid, Spain\\
$^{\boldsymbol{6}}$ Department of Terrestrial Magnetism, Carnegie Institution for Science, 5241 Broad Branch Road NW, Washington DC 20015, USA\\
$^{\boldsymbol{7}}$ Institut de Ci\`encies de l'Espai, Campus UAB, C/Can Magrans s/n, 08193 Bellaterra, Spain\\
$^{\boldsymbol{8}}$ Institut d'Estudis Espacials de Catalunya (IEEC), 08034 Barcelona, Spain\\
$^{\boldsymbol{9}}$ School of Physics and Astronomy, Queen Mary University of London, 327 Mile End Rd, E14NS London, United Kingdom\\
$^{\boldsymbol{10}}$ Institut f\"ur Astrophysik, Georg-August-Universit\"at, Friedrich-Hund-Platz 1, 37077, G\"ottingen, Germany\\
$^{\boldsymbol{11}}$ Centro de Astrobiolog\'ia, CSIC-INTA, ESAC campus, Camino bajo del castillo s/n, 28692, Villanueva de la Ca\~nada, Madrid, Spain\\
$^{\boldsymbol{12}}$ Landessternwarte, Zentrum f\"ur Astronomie der Universit\"at Heidelberg, K\"onigstuhl 12, 69117 Heidelberg, Germany\\
$^{\boldsymbol{13}}$ UCO/Lick Observatory, University of California at Santa Cruz, 1156 High Street, Santa Cruz,
CA 95064, USA\\
$^{\boldsymbol{14}}$ The Observatories, Carnegie Institution for Science, 813 Santa Barbara Street, Pasadena, CA,
91101, USA\\
$^{\boldsymbol{15}}$ Universidad de Chile, Departmento de Astronom\'ia,
Camino El Observatorio 1515, Las Condes, Santiago, Chile\\
$^{\boldsymbol{16}}$ Kavli Institute, Massachusetts Institute of Technology, 77 Massachusetts Avenue, Cambridge,
MA 02139, USA\\
$^{\boldsymbol{17}}$ Department of Exploitation and Exploration of Mines, University of Oviedo, E-33004, Oviedo, Spain\\
$^{\boldsymbol{18}}$ Centre for Astrophysics Research, School of Physics, Astronomy and Mathematics, University of Hertfordshire, College Lane, AL10 9AB, Hatfield, UK\\
$^{\boldsymbol{19}}$ Warsaw University Observatory, Aleje Ujazdowskie 4, 00-478 Warszawa, Poland\\
$^{\boldsymbol{20}}$ Department of Astrophysics and Planetary Science, Villanova University, Villanova, Pennsylvania 19085, USA\\
$^{\boldsymbol{21}}$ Centro Astron\'omico Hispano-Alem\'an (CSIC-MPG), Observatorio Astron\'omico de Calar Alto, Sierra de los Filabres, 04550 G\'ergal, Almer\'ia, Spain\\
$^{\boldsymbol{22}}$ Departamento de F\'isica de la Tierra y Astrof\'isica and IPARCOS-UCM (Instituto de F\'isica de Part\'iculas y del Cosmos de la UCM), Facultad CC. F\'isicas, Universidad Complutense de Madrid, E-28040, Madrid, Spain\\
$^{\boldsymbol{23}}$ The American Association of Variable Star Observers, 49 Bay State Road, Cambridge, MA 02138, USA\\
$^{\boldsymbol{24}}$ AstroLAB IRIS, Provinciaal Domein "De Palingbeek", Verbrandemolenstraat 5, B-8902 Zillebeke, Ieper, Belgium\\
$^{\boldsymbol{25}}$ Vereniging Voor Sterrenkunde (VVS), Oude Bleken 12, 2400 Mol, Belgium








\label{lastpage}
\end{document}